\documentclass[showpacs,superscriptaddress,twocolumn,aps,amsmath,amssymb,prb]{revtex4}
\usepackage{bm}
\usepackage{epsfig}
\newcommand{\App}[1]{Appendix\,\ref{#1}}

\newcommand{\nl}{\nonumber \\}

\newcommand{\be}{\begin{equation}}
\newcommand{\ee}{\end{equation}}
\newcommand{\bea}{\begin{eqnarray}}
\newcommand{\eea}{\end{eqnarray}}
\newcommand{\bsube}{\begin{subequations}}
\newcommand{\esube}{\end{subequations}}
\newcommand{\Fig}[1]{Fig.\,\ref{#1}}
\newcommand{\Eq}[1]{Eq.\,(\ref{#1})}
\newcommand{\Eqs}[1]{Eqs.\,(\ref{#1})}

\newcommand{\B}{\text{\scriptsize bath}}
\newcommand{\ti}{\tilde}
\newcommand{\T}{{\rm total}}
\newcommand{\la}{\langle}
\newcommand{\ra}{\rangle}

\newcommand{\bfL}  {\mbox{\boldmath${\cal L}$}}
\newcommand{\bfV}  {\mbox{\boldmath${\cal V}$}}
\newcommand{\bfG}  {\mbox{\boldmath${\cal G}$}}
\newcommand{\bfone}  {\mbox{\boldmath${\cal I}$}}
\newcommand{\bfS}  {\mbox{\boldmath${\cal S}$}}

\begin{document}

\title{Inelastic transport dynamics through attractive impurity in charge-Kondo regime}

\author{Feng Jiang}\email{fjiang@shiep.edu.cn}
\affiliation{Department of Mathematics and Physics, Shanghai University of
Electric Power, Shanghai, 200090, China}

\author{Shikuan Wang}
\affiliation{Department of Chemistry, Hong Kong University of
Science and Technology, Kowloon, Hong Kong, China}

\author{Hang Xie}
\affiliation{Department of Chemistry, The University of Hong Kong, Pofkulam Road, Hong Kong, China}

\author{Yonghong Yan}
\affiliation{Department of Physics, Shaoxing University, Shaoxing 312000, China}

\author{YiJing Yan}\email{yyan@ust.hk}
\affiliation{Department of Chemistry, Hong Kong University of
Science and Technology, Kowloon, Hong Kong}
\affiliation{Hefei National Laboratory for Physical Sciences at the Microscale,
University of Science and Technology of China, Hefei, Anhui 230026,
China}

\begin{abstract}
Within the frame of quantum dissipation theory, we develop a new hierarchical equations of motion theory, combined with the small polaron transformation. We fully investigate the electron transport of a single attractive impurity system with the strong electron-phonon coupling in charge-Kondo regime.  Numerical results demonstrate the following facts: (i) The density of states curve shows that the attraction mechanism results in not only the charge-Kondo resonance (i.e. elastic pair-transition resonance), but also the inelastic pair-transition resonances and inelastic cotunneling resonances. These signals are separated discernibly in asymmetric levels about Fermi level; (ii) The differential conductance spectrum shows the distinct peaks or wiggles and the abnormal split of pair-transition peaks under the asymmetric bias;  (iii) The improvement of bath-temperature can enhance the phonon-emission (absorption)-assisted sequential tunneling and also strengthen the signals of inelastic pair-transition under the asymmetric bias; (iv) The inelastic dynamics driven by the  ramp-up time-dependent voltage presents clear steps which can be tailored by the duration-time and bath-temperature; (v) The linear-response spectrum obtained from the linear-response theory in the Liouville space reveals the excitation signals of electrons' dynamical transition. Briefly, the physics caused by attraction mechanism stems from the double-occupancy or vacant-occupation of impurity system.
\end{abstract}

\date{\today}

\maketitle

\section{Introduction}
\label{thintro}
In electron transport at nano-scale,
Kondo peaks at the Fermi levels of electrodes often appear at low temperature,
as an interplay between
strong Coulomb interaction within magnetic impurity quantum dot (QD) system
and its coupling with itinerary electrons
from electrodes.\cite{Gol98156,Cro98540}
Most of the work has focused on
steady-state current-voltage characteristics in the strong e-e
Coulomb repulsion regime.\cite{Sch9418436,Sch94423,Kon9531,Kon961715,Che05165324,Wan07045318}
%%%%%%
The resulting Fermi level's resonance has been well understood
as the spin-Kondo effect.\cite{Kon70183} It involves
spin unpaired single-electron occupancy states
of magnetic impurity QD lying below the Fermi energy
of electrodes with the spin paired double-occupancy states far above due to the repulsive Coulomb interaction ($U> 0$).\cite{Mei922512,Mei932601,Hal78416,Pus0145328}
However, the contest between electron-electron (e-e) interaction and
electron-phonon (e-ph) interaction due to the vibration of impurity usually plays a crucial
role in electron transport at nano-scale,
where current-voltage characteristics are often
accompanied by inelastic Franck-Condon steps
of phonon resonances.
Especially, effective Coulomb interaction could become
attractive ($U<0$) when e-ph interaction prevails,
which would lead to the spin paired double-occupancy energy
lying below that of spin single-occupancy state.
\cite{Kuo02085311,Ale03075301,Ale03235312}
Then, the
charge-Kondo effect would take
place instead of the spin-Kondo one.\cite{Cor04147201,Arr05041301,Koc06056803,Koc07195402,And11241107}
Now the double-occupancy is favored so that pair-transition
distinguished to cotunneling
becomes another important
co-operative way of transport.
Simple negative-$U$ models, without an explicit inclusion
of e-ph coupling, or only considering e-ph coupling in zero-bias limitation have been used to
account for the underlying non-equilibrium steady-state
charge-Kondo transport mechanism.\cite{Cor04147201,Arr05041301,Koc06056803,Koc07195402,And11241107}

In this work, we
try to fully investigate inelastic charge-Kondo transport covering finite bias and linear-response region
with negative-$U$ considered using hierarchical equations of motion (HEOM)
method combined with small polaron transformation developed in the former works \cite{Jin08234703,Zhe08184112,Zhe08093016,Zhe09124508,Zhe09164708,Jia12245427} within the frame of quantum dissipation theory and on the basis of adiabatic approximation (the approximation in second-order cumulant expansion). In all the theoretical tools to deal with quantum transport, HEOM based on small polaron transformation has unique advantages such as clear structure, highly efficient numerical achievement, non-perturbative and conveniently time-dependent handling. HEOM has the advantage in both numerical efficiency and applications to various
systems.\cite{Zhe08184112,Zhe08093016,Zhe09124508,Zhe09164708,Tan89101,Tan906676,Tan06082001,%
Xu05041103,Xu07031107,Jin07134113,Zhu115678,Xu11497,Din11164107,Jia12245427}
Moreover, the initial system--environment coupling that is not
contained in the original path integral formalism can now be
accounted for via proper initial conditions to HEOM.

This paper is organized as follows.
In section \textrm{II}, we introduce
our model hamiltonian and describe the calculation method. In section \textrm{III},
firstly, the physical processes induced by the negative-$U$ model are summarized. Then,
the mechanism of elastic pair-transition is analyzed by elastic spectrum function of asymmetric levels without considering e-ph coupling,
further, the mechanism of inelastic pair-transition and inelastic cotunneling are revealed by inelastic spectrum function of the same system with e-ph coupling considered explicitly. On the basis of the differential
conductances under asymmetric bias with considering the symmetric system-electrode coupling and the asymmetric one, the effect of bath-temperature enhancement are analyzed.
In section \textrm{IV}, in conjunction with steady transport, the results and analysis are
given for the inelastic dynamics and the inelastic linear-response spectrums. Finally, we conclude in section \textrm{V}.
\App{thapp_phlr} builds the transformation from
the negative-\emph{U} model to the positive-\emph{U} model.

\section{Inelastic quantum transport: An impurity model and method}
\label{thsec2}

%\subsection{An inelastic quantum transport model based on small polaron transformation}
\subsection{The model}
\label{thsec2A}

Supposing a single impurity QD coupled with a phonon bath is sandwiched by
two electrodes, the e-ph coupling
is canceled by small polaron transformation, \cite{Jia12245427} and the
total hamiltonian is transformed as follows:
\begin{align}
H_{\rm T}&=H_{\rm res}+H_{\rm e}+H_{\rm e-res}+H_{\rm ph},\nl
H_{\rm res}&=\sum_{\alpha,s,k}(\epsilon_{\alpha,s,k}-eV_{\alpha})\hat{c}^{\dag}_{\alpha,s,k}\hat{c}_{\alpha,s,k},\nl
H_{\rm e}&=\sum_{s}\varepsilon_{s}\hat{a}^{\dag}_{s}\hat{a}_{s}+U\hat{a}^{\dag}_{\uparrow}\hat{a}_{\uparrow}\hat{a}^{\dag}_{\downarrow}\hat{a}_{\downarrow},\nl
H_{\rm e-res}&=\sum_{\alpha,s,k}t_{\alpha,s,k}\hat{c}^{\dag}_{\alpha,s,k}\hat{a}_{s}e^{i\hat{\varphi}}+{\rm
H.c.},\nl
H_{\rm ph}&=\sum_{q}\omega_{q}\hat{d}^{\dag}_{q}\hat{d}_{q},
\end{align}
where, $\epsilon_{\alpha,s,k}$ is the $k^{th}$ level of spin $s$ of $\alpha$-electrode in equilibrium, $V_{\alpha}$ ($V_{L}=aV=a(V_{L}-V_{R}),0\leq{a}\leq1$) is the electrical potential applied at $\alpha$-electrode, $\varepsilon_{s}=\varepsilon_{0s}-\lambda$ is the level renormalized from the level $\varepsilon_{0s}$ before transformation minusing reorganization energy $\lambda$, $U=U_{0}-2\lambda$ denotes effective interaction between spin up electron and spin down one renormalized from the interaction $U_{0}$ before transformation by plusing $-2\lambda$, $t_{\alpha,s,k}$ describes the coupling between $\alpha$-electrode and quantum dot, $\hat{\varphi}$ is a hermite phonon operator, which comes from small polaron transformation. \cite{Jia12245427}
Not losing the generality, in the calculations we adopt Einstein lattice model, in which all the vibration modes are assumed to have the same frequency $\Omega$.
We make an appointment that the left electrode is the same as the right one so that $\epsilon_{L,s,k}=\epsilon_{R,s,k}$, and
$\epsilon_{\alpha,s,k}=-\epsilon_{\alpha,s,-k}$, $t_{\alpha,s,k}=t^{*}_{\alpha,s,-k}$, after that, $k=0$ denotes fermi level of electrode due to the symmetry between electron and hole. The appointment is favorable to particle-hole/left-right transformation in \App{thapp_phlr}.

When considering inelastic transport, the density of states
(DOS) describes all the equivalent conducting channels, i.e. the main peaks $\varepsilon_{s}$, $\varepsilon_{s}+U$, and the phonon sidebands
around the main peaks, which are divided by phonon frequency $\Omega$. \cite{Lun02075303,Zhu03165326,Che05165324,Wan07045318}
The micro-process reflects the transition between
the four enlarged Fock states:
$|0,m_{0}\rangle=|0\rangle\otimes|m_{0}\rangle$, $|\uparrow,m_{s}\rangle=|\uparrow\rangle\otimes|m_{s}\rangle$, $|\downarrow,m'_{s}\rangle=|\downarrow\rangle\otimes|m'_{s}\rangle$
and $|\uparrow\downarrow,m_{d}\rangle=|\uparrow\downarrow\rangle\otimes|m_{d}\rangle$, where $|m_{0}\rangle$, $|m_{s}\rangle$, $|m'_{s}\rangle$
and $|m_{d}\rangle$ are the phonon states, while, $|0\rangle$ denotes vacant-occupancy state, $|\uparrow(\downarrow)\rangle$ denotes single-occupancy state and $|\uparrow\downarrow\rangle$ denotes
double-occupancy state.
When $U>0$, $\Gamma_{\alpha s}>T_{\alpha}$,
$\varepsilon_{s}<E_{\rm f}$ and $\varepsilon_{s}+U>E_{\rm f}$ (where, $E_{\rm f}$ denotes
the Fermi level in equilibrium), the above hamiltonian can describe spin-Kondo effect, in which Kondo peak locates at $\omega=E_{\rm f}$.
When $U<0$, $\Gamma_{\alpha s}>T_{\alpha}$, $\varepsilon_{s}>E_{\rm f}$ and $\varepsilon_{s}+U<E_{\rm f}$, the above hamiltonian can also describe charge-Kondo effect, in which Kondo peak locates at $\omega=\varepsilon_{\uparrow}+\varepsilon_{\downarrow}+U$.

\subsection{Comments on the methodology}
\label{thsec2B}

\subsubsection{Construction of HEOM}
The dynamics quantities in the HEOM formalism of
inelastic quantum transport are a set of well-defined auxiliary density
operators (ADOs). The reduced density operator after small polaron transformation is set to be
the zeroth-tier ADO of the hierarchy, i.e., $\rho^{\prime}(t)\equiv \mathrm{tr}_{\rm env}[\hat{X}\rho_{\rm tot}(t)\hat{X}^{\dag}]\equiv \rho^{\prime}_{0}(t)$
(where, $\hat{X}$ is the unitary operator for implementing small polaron transformation \cite{Jia12245427}) and it is coupled to
a set of first-tier ADOs $\{\rho^{\prime(1)}_{j}\}$ by differential equation. The
index $j\equiv\{m\sigma\alpha sk\}$ with $\bar{m}=-m$ and $\bar{\sigma}=-\sigma$  specifies the decomposed memory-frequency components
of environment correlation functions, where $m$ denotes the number of phonons emitted or absorbed, $\sigma$ denotes $+$ or $-$, $\alpha$ denotes the specified electrode, $s$ denotes the degree of spin and $k$ denotes the pole coming from spectrum decomposition.

We summarize the final HEOM formalism as follows.
We denote an $n^{\rm th}$-tier ADO as\cite{Jin08234703,Zhe09164708}
\be\label{rhon_def}
  \rho^{\prime(n)}_{\bf j}\equiv \rho^{\prime(n)}_{j_1\cdots j_n};
\qquad \forall\ {j}_{r} \in \{j\}, \ \ j \equiv \{m\sigma\alpha sk\}.
\ee
Its associated $(n\pm 1)^{\rm th}$-tier ADOs are
$\big\{\rho^{\prime(n+1)}_{{\bf j}j}\equiv \rho^{\prime(n+1)}_{j_1\cdots j_nj}\big\}$
and $\big\{\rho^{\prime(n-1)}_{{\bf j}_r}\!
\equiv\rho^{\prime(n-1)}_{j_1\cdots j_{r-1}j_{r+1}\cdots j_n}\big\}$, respectively.
We have
\begin{align}\label{HEOM}
   \dot\rho^{\prime(n)}_{\bf j} =&
   -\big[i{\cal L} + \gamma^{(n)}_{\bf j}(t)\big]\rho^{\prime(n)}_{\bf j}
    -i \sum_{r=1}^{n}(-)^{n-r}\, {\cal C}_{j_r}\, \rho^{\prime(n-1)}_{{\bf j}_r}
\nl &
     -i \sideset{}{'}\sum_{j;\, \sigma,s\in j}\,
      {\cal A}^{\bar\sigma}_{s}\, \rho^{\prime(n+1)}_{{\bf j}j},
\end{align}
with $\gamma^{(0)}=\rho^{\prime(-1)}=0$ for the zero-tier ADO or the
reduced density operator $\rho=\rho^{\prime(0)}$. The last sum runs only
over those $j\neq j_r; r=1,\cdots\!,n$.
In \Eq{HEOM}, $\gamma^{(n)}_{\bf j}(t)$ collects all involving exponents,
\be\label{gamma_n}
 \gamma^{(n)}_{\bf j}(t) = \sum^n_{r=1} \big[
  \gamma_k+im\Omega-\sigma i \Delta_{\alpha}(t) \big]_{m,\sigma,\alpha,s,k\in j_r},
\ee
${\cal A}^{\sigma}_{s}$ and ${\cal C}_j$
are the fermionic superoperators,
defined via their actions on a fermionic/bosonic operator $\hat O$ as
\be \label{calA}
 {\cal A}^{\sigma}_s\hat O = [\hat a_{s}^{\sigma},\hat O]_{\mp}\, ,
\ee
\be\label{calC}
 {\cal C}_{j} \hat O
 =  (A_m\eta^{\sigma}_k)\, \hat a^{\sigma}_{s} \hat O \pm
  (A_{\bar{m}}\eta^{\bar\sigma}_k)^{\ast} \hat O \hat a^{\sigma}_{\nu s}\, .
\ee
In particular, $\rho^{\prime(n-1)}_{{\bf j}_r}$
and $\rho^{\prime(n+1)}_{{\bf j}j}$
in \Eq{HEOM} are both fermionic or bosonic,
when $n$ is even or odd, respectively.

Using zeroth-tier ADO $\rho^{\prime}_{0}(t)$ and first-tier ADOs $\bm{\rho^{\prime}}_{j}(t)$,
electron number and current can be described as follows:
\begin{align}\label{NC}
N_{s}(t)&=\mathrm{tr}[a^{\dag}_{s}a_{s}\rho^{\prime}_{0}(t)],\nl
I_{\alpha s}(t) &= -2e\,{\rm Im}\big\{ {\rm tr}[\hat a_s\rho^{\prime +}_{\alpha s}(t)]\big\}=-2e\sum_{m,k}{\rm Im}\big\{ {\rm tr}[\hat a_s\rho^{\prime +}_{m\alpha sk}(t)]\}.
\end{align}
and when, $\dot\rho^{\prime(n)}_{\bf j}=0$, the ADOs of steady state can be got as an initial condition, the subsequent calculations of transient physical quantities are very readily to be implemented.
Different physical processes can be distinguished by different HEOM tier-truncation.
For example, HEOM 1st-tier truncation only describes sequential tunneling, HEOM 2nd-tier truncation includes cotunneling process and
HEOM higher-tier truncation includes higher-order cotunneling process.
Numerical calculations have demonstrated \cite{Zhe08093016} that when
$\Gamma_{\alpha s}\ll T_{\alpha}$, 1st-tier truncation of HEOM can supply quantitatively description for the elastic transport process ($\Gamma_{\alpha s}$ denotes the coupling between system and $\alpha$-electrode, $T_{\alpha}$ denotes the temperature of $\alpha$-electrode), when $T_{\alpha}<\Gamma_{\alpha s}<5T_{\alpha}$,  quantitatively descriptions require 2nd-tier truncation of HEOM and when $\Gamma_{\alpha s}>5T_{\alpha}$, at least 3rd-tier truncation can achieve quantitatively description.
For the inelastic transport process studied in our job,  despite $\Gamma_{\alpha s}=20T_{\alpha}$,
HEOM 2nd-tier
truncation based on the Einstein lattice model can still
supply at least qualitative description to inelastic quantum transport because e-ph coupling strength is stronger than electrode-system one.

\subsubsection{Linear response theory based on HEOM}
 To highlight the linearity of HEOM, we arrange
the involving ADOs in a column vector,
denoted symbolically as
\be\label{bfrho}
  {\bm\rho^{\prime}}(t)\equiv \big\{\rho^{\prime}(t),\,\rho^{\prime(1)}_{j}\!(t),\,
   \rho^{\prime(2)}_{j_1\!j_2}\!(t),\, \cdots\,\big\}.
\ee
Thus, \Eq{HEOM} can be recast in a matrix-vector form (each element of
the vector in \Eq{bfrho} is a matrix) as follows,
\be
 \dot{\bm\rho^{\prime}}=-i\bfL(t)\bm\rho^{\prime}, \label{heom-mat-vec}
\ee
with the time-dependent hierarchical-space Liouvillian, as inferred
from \Eqs{HEOM}--(\ref{calC}), being of \be\label{dfLsum}
  \bfL(t)= \bfL_s+\delta\mathcal{L}(t){\bfone} + \delta{\bfV}(t)\, .
\ee
It consists not just the time-independent $\bfL_s$
part, but also two time-dependent parts and each of them will
be treated as perturbation at the linear response level soon. \cite{Wan1301}
Specifically, $\delta\mathcal{L}(t){\bfone}$,
with  ${\bfone}$ denoting the unit operator in the hierarchical Liouville space,
is attributed to a time-dependent external field acting on the reduced system,
while $\delta{\bfV}(t)$ is diagonal and due to the time-dependent potentials
$\delta\Delta_{\alpha}(t)$ applied to electrodes.

We may denote $\delta\Delta_{\alpha}(t)=x_{\alpha}\delta\Delta(t)$,
with $0\leq x_{\rm L}\equiv 1+x_{\rm R} \leq 1$; thus
$\delta\Delta(t)=\delta\Delta_{\rm L}(t)-\delta\Delta_{\rm R}(t)$. It
specifies the additional time-dependent bias voltage, on top of the
constant $V=\mu_{\rm L}-\mu_{\rm R}$, applied across the two
reservoirs. As inferred from \Eq{gamma_n}, we have then
\be\label{del_bfL_prime}
  \delta{\bfV}(t)=-{\bfS}\,\delta\Delta(t) ,
\ee
where ${\bfS}\equiv \text{diag}\big\{0,S^{(n)}_{j_1\cdots j_n};n=1,\cdots,L\big\}$,
with
\be\label{Sn}
  S^{(n)}_{j_1\cdots j_n}\equiv \sum_{r=1}^n
   \big(\sigma x_{\alpha}\big)_{\sigma,\alpha\,\in j_r}.
\ee
Note that  $S^{(0)}=0$.

The additivity of \Eq{dfLsum} and the linearity of HEOM lead readily to
the interaction picture of the HEOM dynamics in response to the
time-dependent external disturbance $\delta\bfL(t)=\delta{\mathcal
L}(t)\bfone +\delta{\bfV}(t)$. %%
The initial unperturbed ADOs vector assumes the nonequilibrium
steady-state form of
\be\label{st_ADOs}
 {\bm\rho^{\prime}}^\text{st}(T,V)\equiv \big\{\bar\rho^{\prime},\,
    \bar\rho^{\prime(1)}_{j},\,
   \bar\rho^{\prime(2)}_{j_1\!j_2},\,\cdots\,\big\},
\ee
under given temperature $T$ and constant bias voltage $V$. It is
obtained as the solutions to the linear equations,
$\bfL_s{\bm\rho^{\prime}}^\text{st}(T,V)=0$, subject to the normalization
condition for the reduced density
matrix.\cite{Jin08234703,Zhe08184112,Zhe09164708}
The unperturbed HEOM propagator reads $\bfG_s(t)\equiv
\exp(-i\bfL_st)$. Based on the first-order perturbation theory,
$\delta\bm\rho^{\prime}(t) \equiv \bm\rho^{\prime}(t)- {\bm\rho^{\prime}}^\text{st}(T,V)$ is then
\be\label{del_bfrho}
 \delta\bm\rho^{\prime}(t)=-i\int_{0}^t\! {d}\tau\,
  \bfG_s(t-\tau)\delta{\bfL}(\tau){\bm\rho^{\prime}}^\text{st}(T,V).
\ee

The response magnitude of a local system observable $\hat A$ is
evaluated by the variation in its expectation value, $\delta A(t) =
{\rm Tr}\{\hat A\delta\rho^{\prime}(t)\}$. Apparently, this involves the
zeroth-tier ADO $\delta\rho(t)$ in
$\delta\bm\rho^{\prime}(t)
 \equiv \big\{\delta\rho^{\prime}(t),\,
   \delta\rho^{\prime(n)}_{j_1\cdots j_n}(t);  n=1,\cdots,L\big\}$.
In contrast, the response current under applied voltages cannot be
extracted from $\delta\rho^{\prime}(t)$, because the current operator is not a
local system observable. Instead, as inferred from \Eq{NC}, while
the steady-state current $\bar I_{\alpha}$ through $\alpha$-reservoir
is related to the steady-state first-tier ADOs, $\bar\rho^{\prime(1)}_j\equiv
\bar\rho^{\prime\sigma}_{m\alpha\mu k}$, the response time-dependent current,
$\delta I_{\alpha}(t)=I_{\alpha}(t)-\bar I_{\alpha}$, is obtained via
$\delta\rho^{\prime(1)}_j(t)=\delta \rho^{\prime\sigma}_{m\alpha\mu k}(t)$.

The above two situations will be treated respectively, by considering
$\delta{\bfL}(t)=\delta{\mathcal L}(t)\bfone$ such as charge-gate linear response and
$\delta{\bfL}(t)=\delta\bfV(t)$ such as charge-bias or current-bias linear response.

\subsubsection{Correlation functions of system}
The DOS $A_{s}(\omega)$ is a strong tool to analyze the transport behavior, which involves the calculations of two correlation functions of system such as
$C_{AB}(t-\tau)=\la \hat A(t)\hat
B(\tau) \ra_{\rm st}$.

We now focus on the evaluation of local system correlation/response
functions with the HEOM approach. This is based on the equivalence
between the HEOM-space linear response theory of \Eq{del_bfrho} and
that of the full system-plus-bath composite space.

We start with the evaluation of nonequilibrium steady-state correlation
function $C_{AB}(t)=\la \hat A(t)\hat B(0) \ra_{\rm st}$, as follows.
For e-ph coupling system, by small polaron transformation, the system correlation function can be recast into the
form of
\begin{align}\label{Cab_def}
 C_{AB}(t)&={\rm Tr}_{\T} \big\{\hat A(t)
 \hat B\rho^{\rm st}_{\T}(T_{\rm res},T_{\rm ph},V)\big\}\nl
&={\rm Tr}_{\T} \big\{[\hat{X}\hat A(t)\hat{X}^{\dag}][\hat{X}\hat{B}\hat{X}^{\dag}][\hat{X}{\rho}^{\rm st}_{\T}(T_{\rm res},T_{\rm ph},V)\hat{X}^{\dag}]\big\}\nl
&={\rm Tr}_{\T} \big\{\hat A^{\prime}(t)\hat{B}^{\prime}{\rho}^{\prime \rm st}_{\T}(T_{\rm res},T_{\rm ph},V)\big\},\nl
%&={\rm Tr}_{\T} \big\{\hat A{\cal G}_{\T}(t)
% [\hat B\rho^{\rm st}_{\T}(T,V)]\big\}
%\nl&
%\equiv {\rm Tr}_{\T}[\hat A \ti\rho_{\T}(t)]
%\nl&
% ={\rm Tr}[\hat A \ti\rho(t)].
\end{align}
where, $\hat A^{\prime}(t)=e^{iH_{\rm T}t}(\hat{X}\hat{A}\hat{X}^{\dag})e^{-iH_{\rm T}t}$, and $H_{\rm T}=H_{\rm res}+H_{\rm e}+H_{\rm e-res}+H_{\rm ph}$. Both $\hat{A}$ and $\hat{B}$ can be factorized as the product of electron-operator and phonon-operator after small polaron transformation.

Considering slow phonon-bath and fast electron, adiabatic approximation leads to the factorization of $C_{AB}$ \cite{Gal06045314}as follows:
\begin{align}
&\quad C_{AB}(t)\nl
 &\approx
{\rm Tr}_{\T} \big\{\hat A(t)\hat{B}{\rho}^{\prime \rm st}_{\T}(T_{\rm res},T_{\rm ph},V)\big\}\nl
&\times{\rm Tr}_{\T}\big\{e^{in\hat{\varphi}(t)}e^{im\hat{\varphi}}{\rho}^{\prime \rm st}_{\T}(T_{\rm res},T_{\rm ph},V)\big\},
\end{align}
where, $n(m)$ is integer, which relates to the number of creation and annihilation operations in $\hat{A}$ or $\hat{B}$.

Generally, due to $e^{i\hat{\varphi}}$ operator included in $H_{\rm e-res}$, $[{H}_{\rm ph}, H_{\rm T}]\neq 0$.
However, specifically, for the strong reorganization energy $\lambda$, $e^{i\hat{\varphi}}$ can be replaced by its ensemble average $\langle e^{i\hat{\varphi}}\rangle_{\rm ph}$, which leads to $[{H}_{\rm ph}, H_{\rm T}]=0$.\cite{Che05165324} Then,
\begin{align}
&\quad C_{AB}(t)\nl
 &\approx
{\rm Tr}_{\T} \big\{\hat A(t)\hat{B}{\rho}^{\prime \rm st}_{\T}(T_{\rm res},T_{\rm ph},V)\big\}\nl
&\times{\rm Tr}_{\rm ph}\big\{e^{in\hat{\tilde{\varphi}}(t)}e^{im\hat{\varphi}}{\rho}^{\rm eq}_{\rm ph}(T_{\rm ph})\big\}\nl
&=C^{\rm e}_{AB}C^{\rm ph}_{AB},
\end{align}
where, $\hat{\tilde{\varphi}}(t)=e^{iH_{\rm ph}t}\hat{\varphi}e^{-iH_{\rm ph}t}$.

\begin{align}
C^{\rm e}_{AB}(t)&={\rm Tr}_{\rm total}\big\{\hat A{\cal G}_{\rm total}(t)[{B}{\rho}^{\prime \rm st}_{\T}(T_{\rm res},T_{\rm ph},V)]\big\}\nl
&\equiv{\rm Tr}_{\rm total}\big\{\hat A\ti\rho_{\rm total}(t)\big\}\nl
&={\rm Tr}_{\rm total}\big\{\hat A\ti\rho(t)\big\}.
\end{align}
The $\rho^{\prime \rm st}_{\T}(T_{\rm res},T_{\rm ph},V)$ and ${\cal G}_{\rm total}(t)$ in $C^{\rm e}_{AB}$ are the steady-state electron density operator and the propagator respectively,
in the total system-bath composite space under constant
bias voltage $V$. Define in the last two identities of \Eq{Cab_def} are also
 $\ti\rho_{\rm total}(t)\equiv{\cal G}_{\rm total}(t)\ti\rho_{\rm total}(0)$
and $\ti\rho(t)\equiv {\rm tr}_{\B}\ti\rho_{\rm total}(t)$,
with
%\be\label{tirhoT0}
 $\ti\rho_{\rm total}(0) = \hat B\rho^{\prime \rm st}_{\rm total}(T_{\rm res},T_{\rm ph},V)$.
%\ee
$C^{\rm e}_{AB}$ can be considered in terms of the linear
response theory, in which the perturbation Liouvillian induced by an
external field $\delta\epsilon(t)$ assumes the form of $-i\delta{\cal
L}(t)(\cdot)=\hat B(\cdot)\delta\epsilon(t)$, followed by the
observation on the local system dynamical variable $\hat A$. Both $\hat
A$ and $\hat B$ can be non-Hermitian. Moreover, $\delta{\cal L}(t)$ is
treated formally as a perturbation and can be a one-side action rather
than having a commutator form.

For the evaluation of $C^{\rm e}_{AB}(t)$ with the HEOM-space dynamics, the
corresponding perturbation Liouvillian is $\delta{\bfL}(t)=\delta{\cal
L}(t)\bfone$, with the above defined $\delta{\cal L}(t)$. It leads to
$-i\delta{\bfL}(\tau){\bm\rho}^{\prime\rm st}(T_{\rm res},T_{\rm ph},V) =\hat B {\bm\rho}^{\prime\rm
st}(T_{\rm res},T_{\rm ph},V) \delta\epsilon(\tau)$ involved in \Eq{del_bfrho}. The linear
response theory that leads to the last identity of \Eq{Cab_def} is now
of the $\ti\rho(t)$ being just the zeroth-tier component of
\be\label{tibmrho_t}
 \ti{\bm\rho}(t)\equiv \big\{\ti\rho(t),\,\ti\rho^{(1)}_{j}\!(t),\,
   \ti\rho^{(2)}_{j_1\!j_2}\!(t),\,\cdots\,\big\}
={\bfG}_s(t)\ti{\bm\rho}(0), \ee
with the initial value of [\emph{cf.}~\Eq{st_ADOs}]
\be\label{tibmrho_init1}
 \ti{\bm\rho}(0)=\hat B{\bm\rho}^{\prime\rm st}(T,V)
 = \big\{\hat B\bar\rho^{\prime},\hat B\bar\rho^{\prime(1)}_{j}\!,\,
    \hat B\bar\rho^{\prime(2)}_{j_1\!j_2},\,\cdots\,\big\}\,.
\ee

$C^{\rm ph}_{AB}$ can be tackled very readily because it satisfy the Gaussian statistics and Wick's theorem for thermodynamic average.

Considering the resolve of correlation function of single impurity $A_{s}(\omega)$ as follows:
\begin{align}
A_{s}(\omega)&\equiv\frac{1}{\pi}\mathrm{Re}[\int^{\infty}_{0}dt\langle|\{\hat{a}_{s}(t),\hat{a}^{\dag}_{s}\}|\rangle e^{i\omega t}],\nl
&=\frac{1}{\pi}\mathrm{Re}[\int^{\infty}_{0}dt(\langle|\hat{a}_{s}(t)\hat{a}^{\dag}_{s}|\rangle+\langle|\hat{a}^{\dag}_{s}(t)\hat{a}_{s}|\rangle^{*})e^{i\omega t}],
\end{align}
which can be transformed to the calculations of two correlation functions $\langle|\hat{a}_{s}(t)\hat{a}^{\dag}_{s}|\rangle$ and $\langle|\hat{a}^{\dag}_{s}(t)\hat{a}_{s}|\rangle$. The $C^{\rm ph}_{AB}$ is just phonon correlation function $C_{\rm ph}(t)=\sum^{+\infty}_{m=-\infty}A_{m}e^{-im\Omega t}$ with Einstein lattice model considered.\cite{Jia12245427}

\section{steady inelastic transport}
\label{thsec3}
Inelastic transport through attractive impurity in charge-Kondo regime presents several novel characteristics as follows:

\noindent ({\it i}) The double-occupancy $P_{\uparrow\downarrow}$ or vacant-occupancy $P_{0}$ of QD dominates the transport process. In equilibrium, when $\varepsilon_{\uparrow}+\varepsilon_{\downarrow}+U<0$, $P_{\uparrow\downarrow}>P_{0}$, when $\varepsilon_{\uparrow}+\varepsilon_{\downarrow}+U>0$,
$P_{\uparrow\downarrow}<P_{0}$, and when $\varepsilon_{\uparrow}+\varepsilon_{\downarrow}+U=0$, $P_{\uparrow\downarrow}=P_{0}$ for degenerate levels.

\noindent ({\it ii}) Charge-Kondo peak i.e. elastic pair-transition peak is robust for Zeeman split of QD level (see \App{thapp_phlr}) i.e. it does not split under magnetic field. Inelastic cotunneling and pair-transition sidebands appear. The cotunneling peaks and the pair-transition peaks suffer the reverse split under asymmetric bias (see \Fig{fig3}).

\noindent ({\it iii}) For symmetric bias (including zero bias) and symmetric system-electrode coupling, inelastic charge-Kondo electron transport equivalents the inelastic electron-hole with positive-$U$ repulsion transport of charge-battery, while for asymmetric system-electrode one, it equivalents that of anti-parrel polarization for the left and right ferromagnetic electrodes (see \App{thapp_phlr}).

\noindent ({\it iv}) For asymmetric bias and symmetric system-electrode coupling, inelastic charge-Kondo electron transport equivalents the inelastic electron-hole with positive-$U$ repulsion transport of spin-cell, while for asymmetric system-electrode one, it equivalents that of anti-parrel polarization for the ferromagnetic electrodes of spin-cell (see \App{thapp_phlr}).

\begin{figure}
\includegraphics[width=1.0\columnwidth]{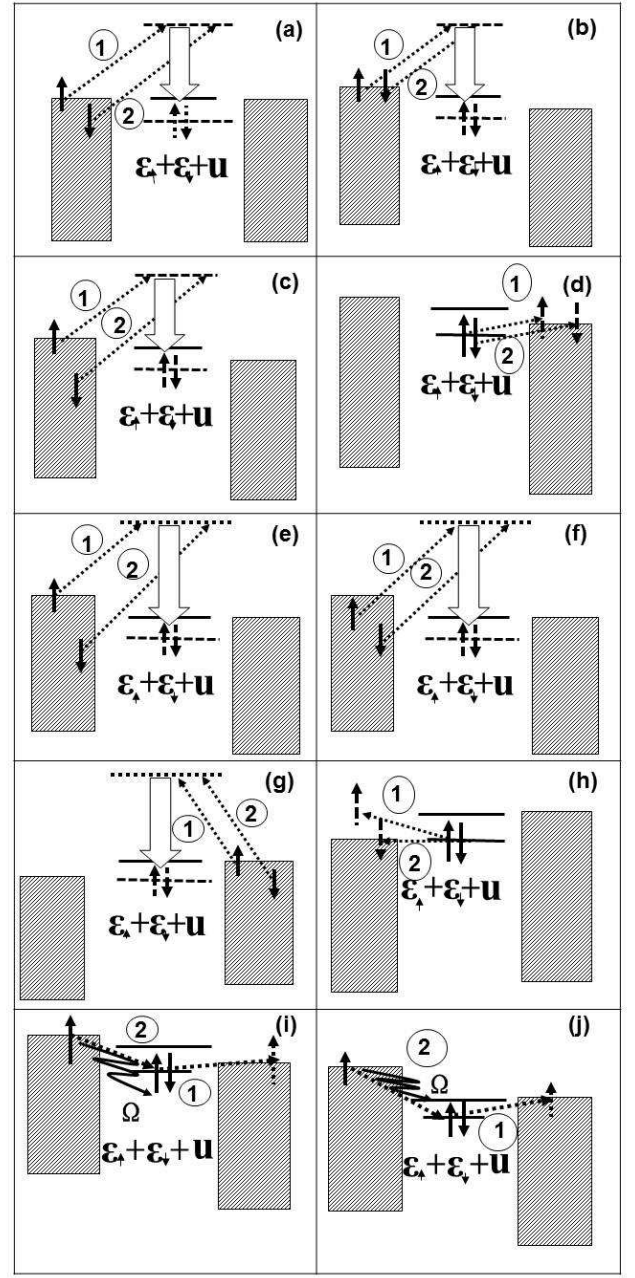}
\caption{The physical processes of charge-Kondo model,
(a) elastic pair-transition process of charge-Kondo model in equilibrium, (b), (c) and (d) elastic pair-transition processes of charge-Kondo model under symmetric bias, (e), (f), (g) and (h) elastic pair-transition processes of charge-Kondo model under positive or negative asymmetric bias, (i) inelastic cotunneling process of charge-Kondo model under symmetric bias, and (j) inelastic cotunneling process of charge-Kondo model under asymmetric bias. The big arrow means the double-occupancy induces the total energy decreasing.}
\label{fig1}
\end{figure}

Considering the symmetry between electron and hole, we only investigate the case of $\varepsilon_{\uparrow}+\varepsilon_{\downarrow}+U<0$ in this paper and the case of $\varepsilon_{\uparrow}+\varepsilon_{\downarrow}+U>0$ can be analyzed similarly. According to \App{thapp_phlr}, the physical processes both in equilibrium and non-equilibrium of charge-Kondo model can be obtained by mapping it to spin-Kondo one via particle-hole/left-right transformation, which are presented clearly in \Fig{fig1} (a)-(j). Only elastic pair-transition and inelastic cotunneling processes are shown and the inelastic pair-transition processes can be achieved only by phonon-emission on the basis of elastic pair-transition. These processes can be explained by the following spectrum function of impurity.

\subsection{The elastic spectrum function of impurity}

\label{thsec2A}
\begin{figure}
\includegraphics[width=1.0\columnwidth]{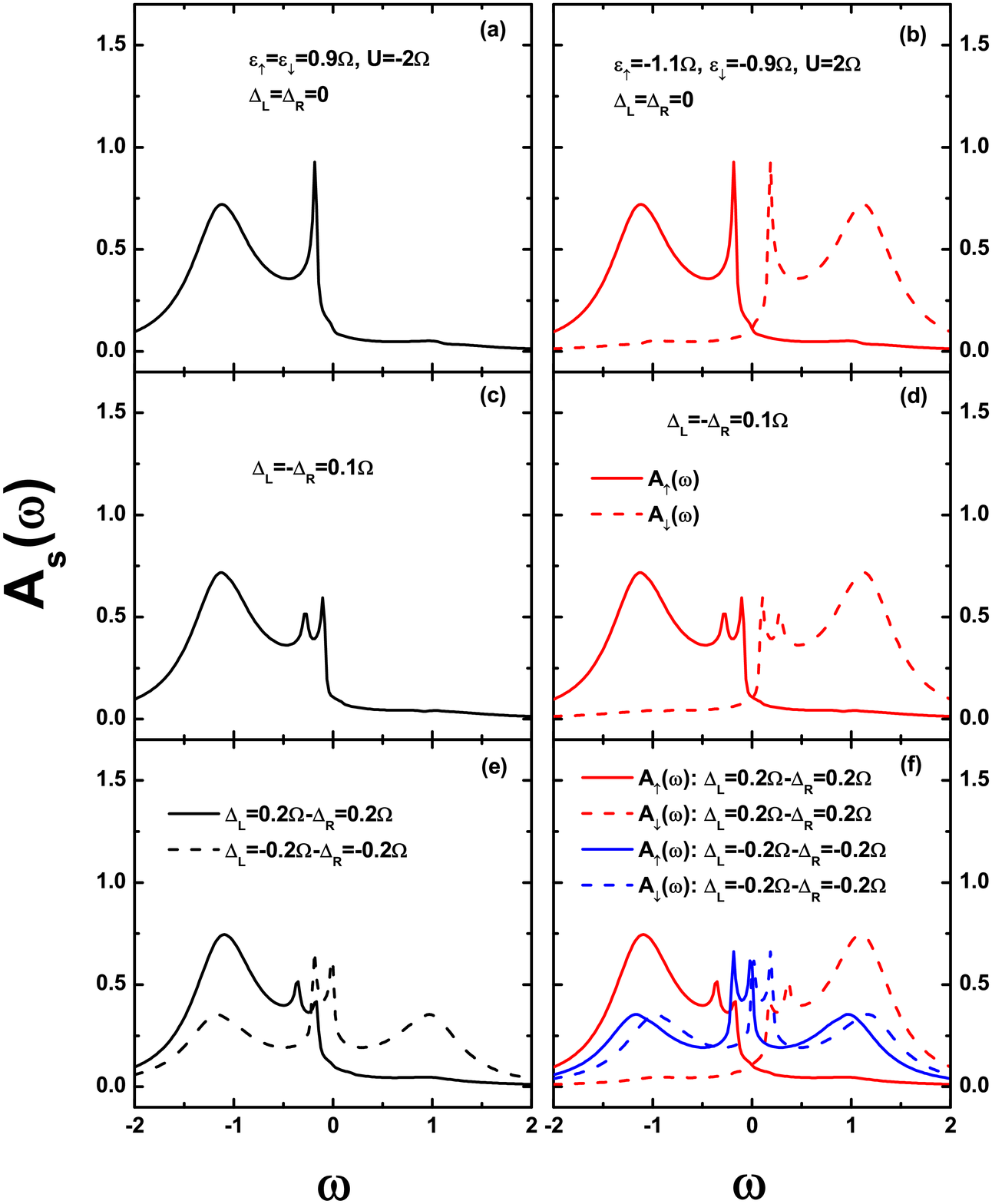}
\caption{The DOS $A_{s}(\omega)$ of spin up (spin down) (in unit of
$\frac{1}{\Omega}$), as function of $\omega$
(in unit of $\Omega$),
for the negative-$U$ Anderson impurity QD system without including e-ph coupling, ${\varepsilon}_{s}=0.9\Omega$, $U=-2\Omega$, (after particle/hole-left/right transformation, ${\varepsilon}_{\uparrow}=-1.1\Omega$, ${\varepsilon}_{\uparrow}=-0.9\Omega$ and $U=2\Omega$),
and $T=T_{\rm res}=T_{\rm ph}=0.01\Omega$:
(a) $\Delta_{\rm L}=\Delta_{\rm R}=0$ and negative-$U$ model, (b) $\Delta_{\rm L}=\Delta_{\rm R}=0$ and positive-$U$ model,
(c) $\Delta_{\rm L}=-\Delta_{\rm R}=0.1\Omega$ and negative-$U$ model, (d) $\Delta_{\rm L}=-\Delta_{\rm R}=0.1\Omega$ and positive-$U$ charge-battery model,
(e) $\Delta_{\rm L}=\pm0.2\Omega-\Delta_{\rm R}=\pm0.2\Omega$ and negative-$U$ model, and (f) $\Delta_{\rm L}=\pm0.2\Omega-\Delta_{\rm R}=\pm0.2\Omega$ and positive-$U$ spin-cell model.
Other parameters are  $\Gamma_{\rm L}=\Gamma_{\rm R}=0.2\Omega$, $W_{\rm L}=W_{\rm R}=30\Omega$.
}
\label{fig2}
\end{figure}

The physical mechanism of charge-Kondo effect can be illustrated clearly by only the simple negative-$U$ model without considering e-ph coupling explicitly.
For the negative-$U$ Anderson impurity QD system, $\varepsilon_{s}=0.9\Omega$, $U=-2\Omega$, $\Gamma_{\rm L}=\Gamma_{\rm L,s}=\Gamma_{\rm R,s}=\Gamma_{\rm R}=0.2\Omega$, $W_{\rm L}=W_{\rm L,s}=W_{\rm R,s}=W_{\rm R}=30\Omega$ and $T_{\rm L}=T_{\rm R}=T_{\rm ph}=0.01\Omega$, where $\Omega$ is phonon energy, $W_{\rm L}$ ($W_{\rm R}$) is bandwidth of the left (right) electrode and $T_{\rm res}$ and $T_{\rm ph}$ are temperatures of the electron bath and the phonon bath respectively.
In equilibrium, $\mu_{\rm L}=\mu_{\rm R}=E_{\rm f}=0$, where, $\mu_{\rm L(\rm R)}$ is the chemical potential of the left or the right electrode.
\Fig{fig2} (a), (c) and (e) show the DOS $A_{s}(\omega)$ of spin up (spin down) both in equilibrium and in non-equilibrium, while \Fig{fig2} (b), (d) and (f) show
the corresponding $A_{s}(\omega)$ of positive-$U$ model by particle-hole/left-right transformation.

\noindent ({\it i}) \Fig{fig2} (a) shows that there are three peaks locating at $\omega=-1.1\Omega$, $\omega=-0.2\Omega$ and $\omega=0.9\Omega$ and one wiggle
locating around $\omega=0$, in which the peaks locating at $\omega=0.9\Omega$ and $\omega=-1.1\Omega$ represent the resonant levels $\varepsilon_{s}=0.9\Omega$
and $\varepsilon_{s}+U=-1.1\Omega$. However, the charge-Kondo peak locating at $\omega=-0.2\Omega$ violates the traditional impression to spin-Kondo one. But, this can be illustrated using particle-hole/left-right transformation, by which, the negative-$U$ model is converted to positive-$U$ one with Zeeman-split ($\varepsilon_{\uparrow}=-1.1\Omega$ and $\varepsilon_{\downarrow}=-0.9\Omega$) shown in \Fig{fig2} (b). With Zeeman-split, spin-Kondo peak also splits to double peaks locating at $\omega=-|\varepsilon_{\uparrow}-\varepsilon_{\downarrow}|$ for spin-up and $\omega=|\varepsilon_{\uparrow}-\varepsilon_{\downarrow}|$ for spin-down, and the wiggle around $\omega=0$ also appears. In spin-Kondo model, spin-flip cotunneling process dominates the transport which is reflected that the peak locates at $\omega=-0.2\Omega$ ($\omega=0.2\Omega$) for spin-up (spin-down) and the wiggle locates around $\omega=0$ for spin-down (spin-up).
Considering $A_{\uparrow}(\omega)=\tilde{A}_{\uparrow}(\omega)$ and $A_{\downarrow}(\omega)=\tilde{A}_{\downarrow}(-\omega)$ ($A_{\uparrow(\downarrow)}(\omega)$ is the DOS of negative-$U$ model and $\tilde{A}_{\uparrow(\downarrow)}(\omega)$ is the DOS of positive-$U$ model),
the corresponding pair-transition process achieves, in which spin-down (spin-up) electron on the Fermi level combined with spin-up (spin-down) electron locating at $\omega=-0.2\Omega$ of the left electrode can transit into the QD of vacant-occupancy (refer to \Fig{fig1} (a)),
or the QD of double-occupancy with the total energy $\omega=-0.2\Omega$ transit into the Fermi level and the state with $\omega=-0.2\Omega$ of the right electrode, so the charge-Kondo peak locating at $\omega=-0.2\Omega$ and the wiggle locating around $\omega=0$ appear in \Fig{fig2} (a). But due to Pauli exclusion principle, double-occupancy transition can not happen really in equilibrium.

\noindent ({\it ii}) \Fig{fig2} (c) shows that when symmetric bias $\Delta_{\rm L}=-\Delta_{\rm R}=0.1\Omega$ is applied, the charge-Kondo peak and the wiggle locating at $\omega=0$ split so that
two peaks locate at $\omega=-0.3\Omega$ and $\omega=-0.1\Omega$ with the wiggle locating around $\omega=0.1\Omega$. The positive-$U$ model also shows all the peaks and the wiggle split with $\Delta\omega=\pm 0.1\Omega$ in \Fig{fig2} (d). The signal of spin up locating at $\omega=-0.1\Omega$ ($\omega=-0.3\Omega$ or $\omega=0.1\Omega$) combining the signal of spin down locating at $\omega=0.1\Omega$ ($\omega=-0.1\Omega$ or $\omega=0.3\Omega$) represents a spin-flip cotunneling process under symmetric bias. The corresponding electron-pairs in the left electrode with the reverse spin ($-0.1\Omega\uparrow$, $-0.1\Omega\downarrow$) and ($-0.3\Omega\uparrow(\downarrow)$, $0.1\Omega\downarrow(\uparrow)$) transit into the QD of vacant-occupancy (refer to \Fig{fig1} (b) and (c)), or the double-occupancy electron-pairs in QD transit into vacant states
with the same energy $\epsilon_{\rm k R\uparrow}=\epsilon_{\rm k R\downarrow}=-0.1\Omega$ (refer to \Fig{fig1} (d)), which can happen really in non-equilibrium.

\noindent ({\it iii}) \Fig{fig2} (e) shows that when positive asymmetric bias $\Delta_{\rm L}=0.2\Omega$ is applied, the charge-Kondo peak splits to double peaks locating at $\omega=-0.2\Omega$ and $\omega=-0.4\Omega$, while the wiggle also splits to double wiggles locating at $\omega=0$ and $\omega=0.2\Omega$. \Fig{fig2} (f) shows the spin-flip cotunneling process of positive-$U$ QD sandwiched by spin-cells. Considering Pauli exclusion principle, it can be mapped into electron-pairs in the left electrode with the reverse spin ($-0.4\Omega\uparrow(\downarrow)$, $0.2\Omega\downarrow(\uparrow)$) and ($-0.2\Omega\uparrow(\downarrow)$, $0\downarrow(\uparrow)$) transiting into the QD of vacant-occupancy (refer to \Fig{fig1} (e) and (f)). When negative asymmetric bias $\Delta_{\rm L}=-0.2\Omega$ is applied, the charge-Kondo peak splits to double peaks locating at $\omega=-0.2\Omega$ and $\omega=0$, while the wiggle also splits to double wiggles locating at $\omega=0$ and $\omega=-0.2\Omega$, so because of superposition effect, only two peaks locate at $\omega=-0.2\Omega$ and $\omega=0$ respectively, at the same time, the collapsing main peak locating at $\omega=0.9\Omega$ resumes. Mapping the spin-flip process in positive-$U$ model to pair-transition process in negative-$U$ model, electron-pairs in the right electrode with the reverse spin ($-0.2\Omega\uparrow(\downarrow)$, $0\downarrow(\uparrow)$) transiting into the QD of vacant-occupancy (refer to \Fig{fig1} (g)), or the double-occupancy electron-pair in QD will transit into the vacant-occupancy states with the different energy $\epsilon_{\rm k L\uparrow(\downarrow)}=-0.2\Omega$ and $\epsilon_{\rm k L\downarrow(\uparrow)}=0$ (refer to \Fig{fig1} (h)).

In summary, the DOS presented above does not show the phonon-sidebands because e-ph coupling is not included in hamiltonian, so all the physical processes keep energy conservation. Once considering e-ph coupling, inelastic processes will go into the transport picture and the corresponding DOS is presented in the next section.

\subsection{The inelastic spectrum function of impurity}
\label{thsec2B}
\begin{figure}
\includegraphics[width=1.0\columnwidth]{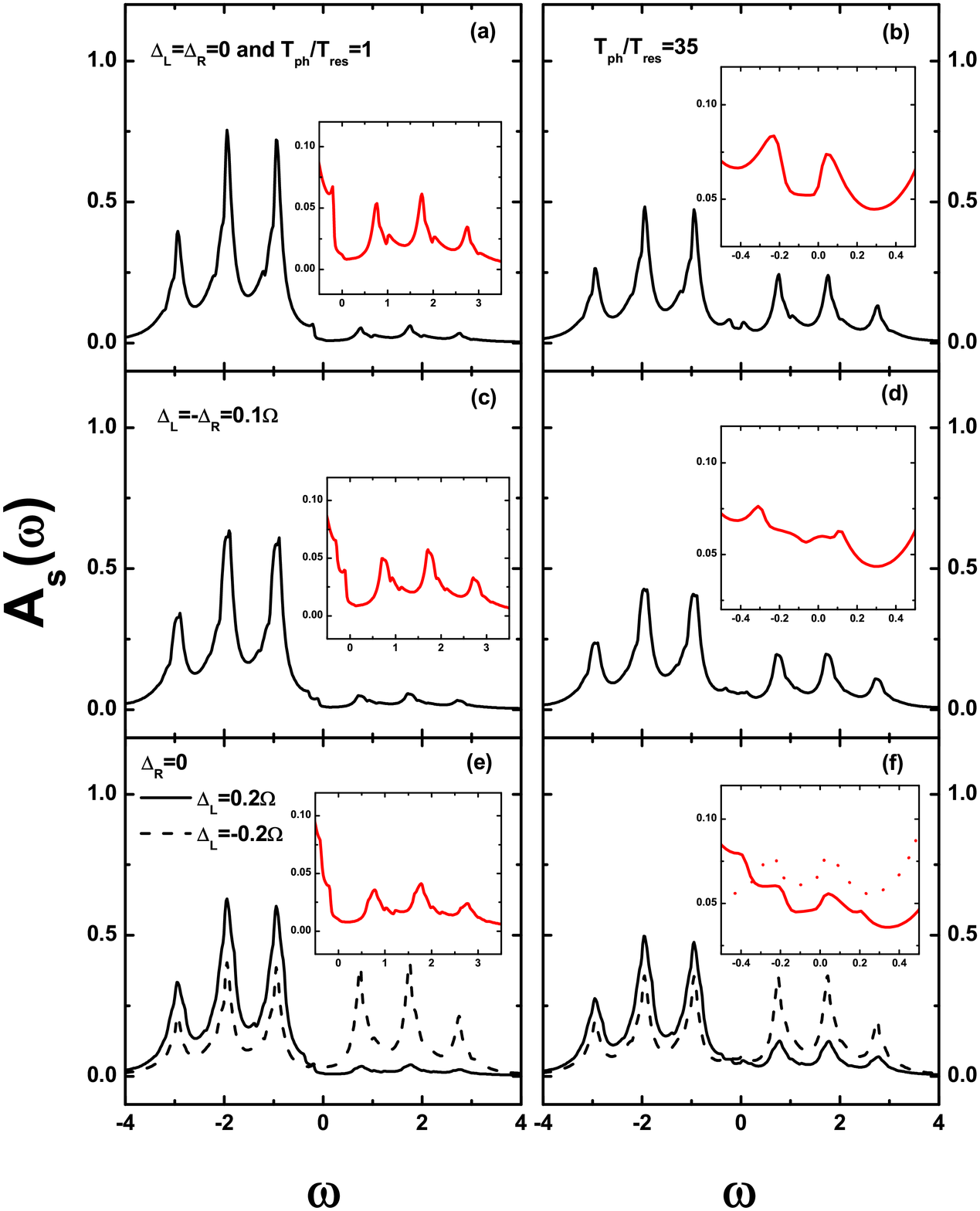}
\caption{The DOS $A_{s}(\omega)$ of spin up (spin down) (in unit of
$\frac{1}{\Omega}$), as function of $\omega$
(in unit of $\Omega$),
for the phonon-coupled Anderson impurity QD system, with ${\varepsilon}_{s}=0.9\Omega$,
$U=-2\Omega$, $T_{\rm res}=0.01\Omega$:
(a) $T_{\rm ph}=0.01\Omega$ and $\Delta_{\rm L}=\Delta_{\rm R}=0$, (b) $T_{\rm ph}=0.35\Omega$ and $\Delta_{\rm L}=\Delta_{\rm R}=0$,
(c) $T_{\rm ph}=0.01\Omega$ and $\Delta_{\rm L}=-\Delta_{\rm R}=0.1\Omega$, (d) $T_{\rm ph}=0.35\Omega$ and $\Delta_{\rm L}=-\Delta_{\rm R}=0.1\Omega$.
(e) $T_{\rm ph}=0.01\Omega$ and $\Delta_{\rm L}=\pm0.2\Omega-\Delta_{\rm R}=\pm0.2\Omega$,
and (f) $T_{\rm ph}=0.35\Omega$ and $\Delta_{\rm L}=\pm0.2\Omega-\Delta_{\rm R}=\pm0.2\Omega$.
Other parameters are the same as those adopted in \Fig{fig2}. The insets
magnify the DOS around $\omega=0$.
}
\label{fig3}
\end{figure}

Strong e-ph coupling will weaken charge-Kondo peak greatly. For the same system, when e-ph coupling is absorbed to hamiltonian, inelastic spectrum functions $A_{s}(\omega)$ are presented in \Fig{fig3} (a)-(f).

\noindent ({\it i}) \Fig{fig3} (a) and the inset show DOS in equilibrium has three series of signals: the first includes the signals of sequential tunneling locating at $\omega=\varepsilon_{s}+m\Omega$ and $\omega=\varepsilon_{s}+U-m\Omega$ ($m=0, 1, 2, \cdots $), the second includes the signals of inelastic cotunneling locating at $\omega=m\Omega$ ($m=\pm1, \pm2, \cdots $), the third includes the signals of pair-transition locating at $\omega=0$ and $\omega=-0.2\Omega+m\Omega$ ($m=0, \pm1, \pm2, \cdots $). \Fig{fig3} (b) and the inset show that when bath-temperature is raised to $T_{\rm ph}=0.35\Omega$, two additional peaks appear with one locating around $\omega=-0.25\Omega$ and the other locating around $\omega=0.05\Omega$. Noting that the peak locating around $\omega=-0.25\Omega$ represents the peak of phonon-absorption and
the peak locating around $\omega=0.05\Omega$ represents the peak of phonon-emission. The elastic pair-transition signals are smoothed.

\noindent ({\it ii}) \Fig{fig3} (c) and the inset show under symmetric bias, the signals of pair-transition and that of inelastic cotunneling split with
$\delta\omega=\pm\mu_{L}$. Inelastic cotunneling denotes the double-side unidirectional pair-transition by one-phonon-emission or multiple-phonon-emission with spin conservation (refer to \Fig{fig1} (i)).
Under the non-equilibrium condition, the wiggle locating at $\omega=0.1\Omega$ coming from the split of the wiggle locating at $\omega=0$ disappears, but it will appear again distinctly with the bath-temperature raised.
\Fig{fig3} (d) and the inset show that under the symmetric bias, the two signals induced by bath-temperature enhancement don't split because they represent sequential tunneling. Another two peaks locating at $\omega=-0.3\Omega$ and $\omega=0.1\Omega$ come from the split of charge-Kondo peak and the wiggle around $\omega=0$, which are covered by the peaks due to hot phonon effect in equilibrium.

\noindent ({\it iii}) \Fig{fig3} (e) and the inset show under asymmetric bias, the signals locating at $\omega=-0.2\Omega+m\Omega$ ($m=0, \pm1, \pm2, \cdots $)
split with one peak locating at the original place and the other shifts with $\delta\omega=-\mu_{L}$, while the signals locating at $\omega=m\Omega$ ($m=0, \pm1, \pm2, \cdots $) split with one peak fixed and the other shifts with $\delta\omega=\mu_{L}$.
In comparison with the signals of inelastic cotunneling, the pair-transition signals have the opposite shift because it contributes to keeping pair-transition both for elastic scattering and for inelastic scattering, the signals of inelastic cotunneling respond to $\Delta_{\rm L}$ also helps to achieve cotunneling by phonon-emission (refer to \Fig{fig1} (j)).
\Fig{fig3} (f) and the inset show that under the asymmetric bias, the two peaks induced by bath-temperature enhancement also keeps no splitting except the splitting of pair-transition peaks and cotunneling peaks.

\subsection{The differential-conductance spectrum under asymmetric bias}
\label{thsec2C}
\begin{figure}
\includegraphics[width=1.0\columnwidth]{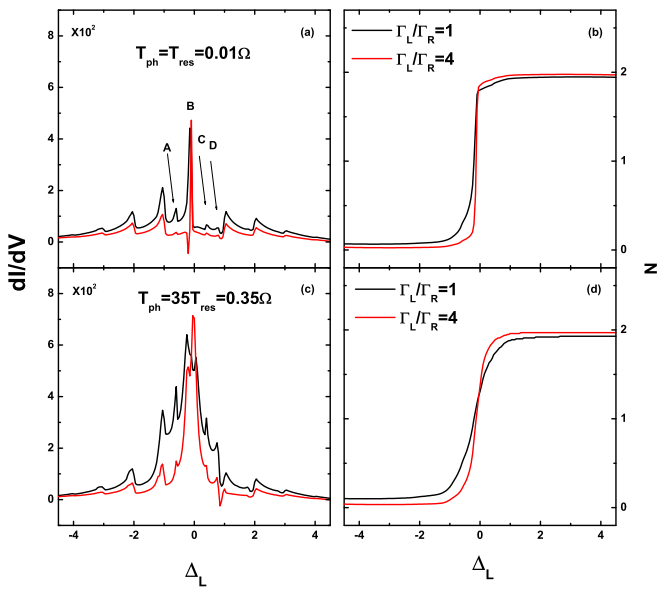}
\caption{The differential-conductance spectrum ${\rm d}I/{\rm d}V$
(in unit of
$\frac{2e^{2}\Gamma_{\rm L}\Gamma_{\rm R}}{\hbar\Omega(\Gamma_{\rm L}+\Gamma_{\rm R})}$),
and electron-number $N$ as function of $\Delta_{\rm L}$ (in unit of $\Omega$) under asymmetric bias for the phonon-coupled Anderson impurity QD system, with ${\varepsilon}_{s}=0.9\Omega$,
$U=-2\Omega$:
(a) $dI/dV$ for $T_{\rm ph}=T_{\rm res}=0.01\Omega$,
(b) $N$-$V$ for $T_{\rm ph}=T_{\rm res}=0.01\Omega$,
(c) $dI/dV$ for $T_{\rm ph}=35T_{\rm res}=0.35\Omega$,
(d) $N$-$V$ for $T_{\rm ph}=35T_{\rm res}=0.35\Omega$.
Other parameters are the same as those adopted in \Fig{fig2}. The black solid line describes $dI/dV$ of
symmetric system-electrode coupling ($\Gamma_{\rm L}=\Gamma_{\rm R}=0.2\Omega$) and the red solid line describes that of asymmetric one ($\Gamma_{\rm L}=4\Gamma_{\rm R}=0.2\Omega$).
}
\label{fig4}
\end{figure}

In conjunction with DOS, differential-conductance $dI/dV$ spectrum discovers the inelastic signals by peaks or steps explicitly and particle-number $N$-$V$ curve helps to reveal the generation and decomposition of small polaron under bias clearly. For charge-Kondo transport, the existing works focus on symmetric bias, \cite{Cor04147201,Arr05041301,Koc06056803,Koc07195402,And11241107} and under symmetric bias, sequential tunneling dominates the transport process. In order to investigate the transport beyond sequential tunneling, we focus on asymmetric bias.
\Fig{fig4} (a)-(d) present $dI/dV$ and $N$-$V$ under asymmetric bias for the different bath-temperature. For the single-side bias, we consider both the case of symmetric system-electrode coupling ($\Gamma_{\rm L}=\Gamma_{\rm R}=0.2\Omega$) and that of asymmetric system-electrode one ($\Gamma_{\rm L}=4\Gamma_{\rm R}=0.2\Omega$).

\Fig{fig4} (a) shows the $dI/dV$ spectrum with $T_{\rm ph}=T_{\rm res}=0.01\Omega$ under asymmetric bias. With asymmetric bias applied,
asymmetric levels induces the asymmetry of current-bias: $I_{\rm L}(\Delta_{\rm L}>0)<-I_{\rm L}(-\Delta_{\rm L})$. $dI/dV$ spectrum shows that
besides the inelastic cotunneling signals locating around $\Delta_{\rm L}=\pm m\Omega$, ($m=1, 2, 3, \cdots$),
there are still four peaks named $\mathbf{A}$, $\mathbf{B}$, $\mathbf{C}$ and $\mathbf{D}$ locating around $\Delta_{\rm L}=-0.6\Omega$, $\Delta_{\rm L}=-0.1\Omega$, $\Delta_{\rm L}=0.4\Omega$ and $\Delta_{\rm L}=0.8\Omega$ respectively.
The peak $\mathbf{A}$ as an inelastic pair-transition signal reflects the process of double-occupancy electron-pair in QD transiting into the vacant levels with the energy $\epsilon_{\rm Lk\uparrow}=\epsilon_{\rm Lk\downarrow}=-0.6\Omega$ by emitting a phonon. The peak $\mathbf{B}$ as an elastic pair-transition signal reflects the process of double-occupancy electron-pair in QD transiting into the vacant levels with the energy $\epsilon_{\rm Lk\uparrow}=\epsilon_{\rm Lk\downarrow}=-0.1\Omega$, keeping energy conservation. The peak $\mathbf{C}$ reflects the process of the electrons achieving inelastic pair-transition from the occupied levels with the energy $\epsilon_{\rm Lk\uparrow}=\epsilon_{\rm Lk\downarrow}=0.4\Omega$ to unoccupied QD by one-phonon--emission.
As another type of inelastic pair-transition, the peak $\mathbf{D}$
reflects the process of the electron-pair with the energy $\epsilon_{\rm Lk\uparrow(\downarrow)}=0.8\Omega$ and $\epsilon_{\rm Lk\downarrow(\uparrow)}=0$
transiting into vacant-occupancy QD by one-phonon--emission. With the reverse bias increasing, the asymmetric coupling also makes the subsequent negative differential conductance when $-0.3\Omega<\Delta_{\rm L}<-0.1\Omega$, which corresponds to the slower decreasing of electron number after elastic pair-transition than that with the symmetric coupling. \Fig{fig4} (b) shows the corresponding $N$-$V$ curves, which presents the sharper trend, compared to symmetric bias. With the positive bias increasing, bipolaron forms rapidly and with the reverse bias increasing, polaron is broken up rapidly because of lacking the charges.

\Fig{fig4} (c) shows the $dI/dV$ spectrum with $T_{\rm ph}=35T_{\rm res}=0.35\Omega$ under asymmetric bias.
Different from symmetric bias with bath-temperature raised, generally speaking, asymmetric bias with bath-temperature rising strengthens the signals of inelastic pair-transition. For symmetric coupling, corresponding to the phonon-absorption peak locating at $\omega=-0.25\Omega$ and the phonon-emission peak locating at $\omega=0.05\Omega$, $dI/dV$ spectrum also presents the phonon-absorption signal locating at $\Delta_{\rm L}=-0.25\Omega$ and the phonon-emission signal locating at $\Delta_{\rm L}=0.05\Omega$, and the phonon-absorption signal is the sharpest among all the signals. While for asymmetric coupling, the signal of the phonon-emission locating at $\Delta_{\rm L}=0$ is sharper. With the positive bias increasing, the asymmetric coupling also makes the subsequent negative differential conductance when $0.8\Omega<\Delta_{\rm L}<1\Omega$, which corresponds to the slower increasing of electron number after inelastic pair-transition than that with the symmetric coupling. The outstanding point is that inelastic pair-transition signals enhanced exceed inelastic cotunneling signals with bath-temperature raised, or in other words, phonon-absorption/emission-sequential tunneling drives the inelastic pair-transition processes. The $N$-$V$ curves presented in \Fig{fig4} (d) show that bath-temperature rising suppresses electron-number of QD and smoothes the curves, but obviously, asymmetric coupling induces the electron-number varies more rapidly than the symmetric one.

In summary, under asymmetric bias, the strong coupling induces the sharper
$dI/dV$ spectrum than that with the weak coupling because the cotunneling dominates the transport process under asymmetric bias.

\section{Inelastic dynamics and linear-response spectrum}
Besides the information of steady transport, we also concern the process of building steady state of an open QD with strong e-ph coupling considered, which demands the simulations of dynamics driven by time-dependent voltage.
In conjunction with simulations of dynamics, linear-response spectrum can
fully reveal the dynamical transition's signals of electrons under very small bias and it reflects the inner-attributes of e-ph open system. Using
HEOM, transient transport and steady one can be dealt with at the same level beyond the some simple forms' bias such as a step function voltage \cite{Wei08195316} and the zero-frequency component of linear-response spectrum recovers the steady differential conductance.

\subsection{The inelastic dynamics driven by time-dependent voltage}
\label{thsec3A}
Without loss the generality, considering asymmetric levels: $\varepsilon_{s}=0.9\Omega$ and $\varepsilon_{s}+U=-1.1\Omega$ are coupled asymmetrically to the left and the right electrodes: $\Gamma_{\rm L}=4\Gamma_{\rm R}=0.2\Omega$. A ramp-up voltage $V(t)=V_{\rm L}(t)-V_{\rm R}(t)$ is applied to the left and the right leads at $t=0$, which excites the QD out of equilibrium. The $\alpha$-lead energy levels are shifted due to the voltage: $\Delta_{\alpha}(t)=-eV_{\alpha}(t)$, with $e$ being the elementary charge. Under a ramp-up voltage, $\Delta_{\alpha}(t)$ varies linearly with time until $t=\tau$, and afterwards is kept at a constant amplitude $\Delta$:
\be
\Delta_{\rm L (\rm R)}(t)=\Bigl\{\begin{array}{cl}
\pm4 t/\tau, &{0\leq{t}\leq\tau},
\\
\pm4, &{\tau<t\leq 34.17},
\end{array}
\ee
where, energy-scale is in unit of $\Omega$ and time-scale is in unit of $\frac{\hbar}{(\Gamma_{\rm L}+\Gamma_{\rm R})}$.

\begin{figure}
\includegraphics[width=1.0\columnwidth]{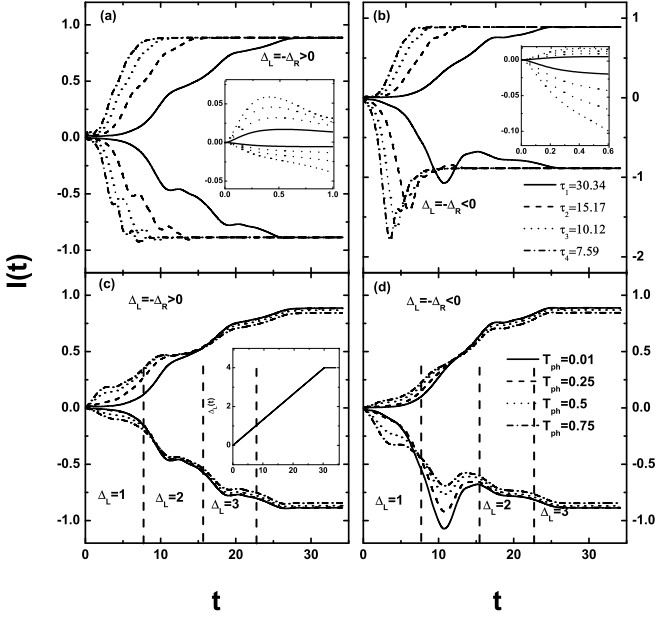}
\caption{The inelastic dynamical evolution $I_{\rm L}>0$ and $I_{\rm R}<0$ (in unit of $\frac{2e\Gamma_{\rm L}\Gamma_{\rm R}}{\hbar(\Gamma_{\rm L}+\Gamma_{\rm R})}$) as function of $t$ (in unit of $\frac{\hbar}{(\Gamma_{\rm L}+\Gamma_{\rm R})}$) driven by ramp-up time-dependent voltage for the phonon-coupled Anderson impurity QD system, with ${\varepsilon}_{s}=0.9\Omega$,
$U=-2\Omega$ and $\Gamma_{\rm L}=4\Gamma_{\rm R}=0.2\Omega$:
(a) under symmetric ramp-up $\Delta_{\rm L}(t)=-\Delta_{\rm R}(t)>0$ and (b) under $\Delta_{\rm L}(t)=-\Delta_{\rm R}(t)<0$ with $T_{\rm ph}=0.01\Omega$ and different duration-time $\tau_{1}=30.34$
(solid line), $\tau_{2}=15.17$ (dash line), $\tau_{3}=10.12$ (dot line) and $\tau_{4}=7.59$ (dash dot line) respectively;
(c) under symmetric ramp-up $\Delta_{\rm L}(t)=-\Delta_{\rm R}(t)>0$ and (d) under $\Delta_{\rm L}(t)=-\Delta_{\rm R}(t)<0$ with duration-time $\tau_{1}=30.34$ and different bath-temperature $T_{\rm ph}=0.01\Omega$ (solid line), $T_{\rm ph}=0.25\Omega$ (dash line), $T_{\rm ph}=0.5\Omega$ (dot line) and $T_{\rm ph}=0.75\Omega$ (dash dot line).
The insets in (a) and (b) magnify the current in very initial time-interval, the inset in (c) presents $\Delta_{\rm L}$ (in unit of $\Omega$) as function of $t$
and the three vertical dash lines in (c) and (d) mark the moments when $\Delta_{\rm L}=1$, $\Delta_{\rm L}=2$ and $\Delta_{\rm L}=3$ respectively.
Other parameters are the same as those adopted in \Fig{fig2}.}
\label{fig5}
\end{figure}

By tuning the duration parameter $\tau$, ramp-up voltage's behavior changes from adiabatic limit to instantaneous switch-on. Here, we explore the intermediate range with a finite $\tau$. \Fig{fig5} (a) and (b) show $I$-$t$ curves under symmetric ramp-up $\Delta_{\rm L}(t)=-\Delta_{\rm R}(t)>0$ and
$\Delta_{\rm L}(t)=-\Delta_{\rm R}(t)<0$ with $T_{\rm res}=T_{\rm ph}=0.01\Omega$. Different duration-time $\tau_{1}=30.34$
(solid line), $\tau_{2}=15.17$ (dash line), $\tau_{3}=10.12$ (dot line) and $\tau_{4}=7.59$ (dash dot line) are considered. The clear steps
appear with time evolution corresponding to main peak resonance, one-phonon-emission and double-phonon-emission respectively. With duration-time deceasing,
the steps will be smoothed and some new satellite peaks will appear, which results from the time-varying phase factor of the non-equilibrium lead correlation function.
With $\tau\rightarrow\infty$, the steps will be very distinct because adiabatic limit corresponds to the steady state.
The insets in (a) and (b) magnify the current in very initial time-interval, which show that the current rises tight followed by a step and hint the initial charge-Kondo resonance. We also observe that when $\Delta_{\rm L}(t)=-\Delta_{\rm R}(t)<0$, $-I_{\rm L}(t)\gg I_{\rm R}(t)$ during a long time-interval which induces that the electron number of QD decreases very rapidly. \Fig{fig5} (c) and (d) show $I$-$t$ curves under symmetric ramp-up $\Delta_{\rm L}(t)=-\Delta_{\rm R}(t)>0$ and
$\Delta_{\rm L}(t)=-\Delta_{\rm R}(t)<0$ with $\tau=30.34$. Different bath-temperature $T_{\rm ph}=0.01\Omega$ (solid line), $T_{\rm ph}=0.25\Omega$ (dash line), $T_{\rm ph}=0.5\Omega$ (dot line) and $T_{\rm ph}=0.75\Omega$ (dash dot line) are considered. The appearing of the first step in $I-t$ curve before the main peak resonance is postponed because local heating induces the broadened one-phonon-emission/absorption peak near Fermi level. It is obvious that the current increases at the same moment with $T_{\rm ph}$ raised initially and decreases at the same moment finally. It is believed that under large bias limitation, the steady current decreases with bath-temperature raised.

\subsection{The current/charge-bias and charge-gate linear-response spectrum}
\label{thsec3B}
If the time impulse $\Delta(t)$ is applied symmetrically at the double leads, the frequency-dispersed dynamical admittance of linear response $G_{\alpha}(\omega)=\frac{\delta I_{\alpha}(\omega)}{\delta V(\omega)}=0.5[\chi_{\alpha L}(\omega)-\chi_{\alpha R}(\omega)]$, where
$\delta I_{\alpha}(\omega)$ and $\delta V(\omega)$ are the Fourier transformation of $\delta I_{\alpha}(t)=I_{\alpha}(t)-I_{\alpha}(0)$ and $\delta V(t)=
(\Delta(t)-\Delta(0))/e$ respectively,
$\chi_{\alpha\alpha^{'}}(\omega)$ is the linear response spectrum of the
$\alpha$-electrode under the asymmetric zero-bias limit applied to the $\alpha'$-electrode.
For the symmetric system-electrode coupling, $\chi_{\rm LL}(\omega)=\chi_{\rm RR}(\omega)$
and $\chi_{\rm LR}(\omega)=\chi_{\rm RL}(\omega)$. For the asymmetric system-electrode coupling, $\chi_{\rm LL}(\omega)\neq\chi_{\rm RR}(\omega)$, but
$\chi_{\rm LR}(\omega)$ still equals to $\chi_{\rm RL}(\omega)$.

\begin{figure*}
\includegraphics[width=1.5\columnwidth]{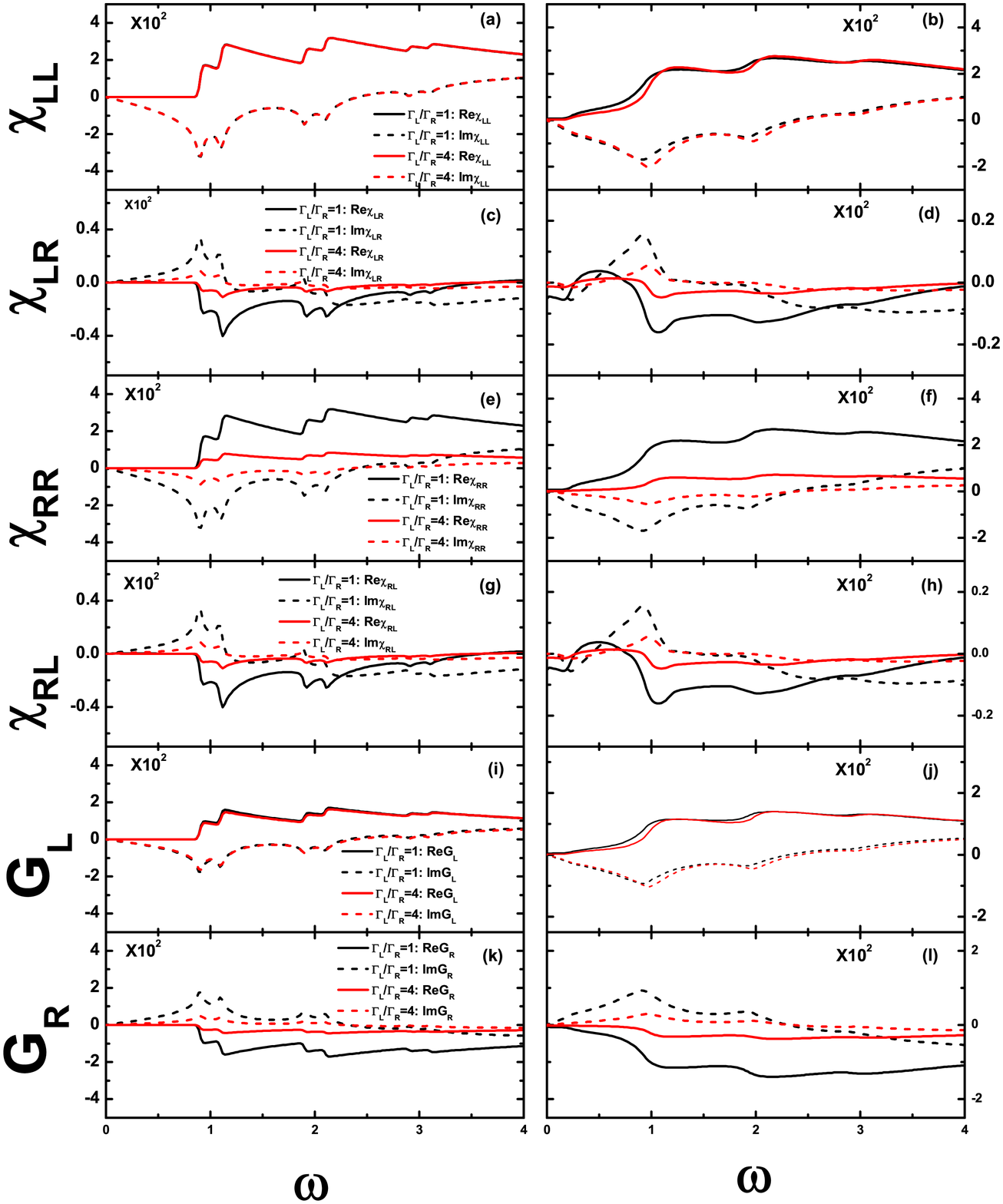}
\caption{The current-bias linear-response spectrum
(in unit of
$\frac{2e^{2}}{\hbar}$),
as function of $\omega$
(in unit of $\Omega$),
for the phonon-coupled Anderson impurity QD system, with ${\varepsilon}_{s}=0.9\Omega$,
$U=-2\Omega$:
(a) $\chi_{\rm L \rm L}(\omega)$, (c) $\chi_{\rm L \rm R}(\omega)$,
(e) $\chi_{\rm R \rm R}(\omega)$, (g) $\chi_{\rm R \rm L}(\omega)$, (i) $G_{\rm L}(\omega)$ and (k) $G_{\rm R}(\omega)$ for HEOM 1st-tier truncation;
(b), (d), (f), (h), (j) and (l) for HEOM 2nd-tier truncation.
The black (red) solid line means
the real part of linear-response function for symmetric (asymmetric) system-electrode coupling: $\Gamma_{\rm L}=\Gamma_{\rm R}=0.2\Omega$ ($\Gamma_{\rm L}=4\Gamma_{\rm R}=0.2\Omega$), while the black (red) dash line means the
imaginary part of linear-response function for symmetric (asymmetric) system-electrode coupling.
Other parameters are the same as that adopted in \Fig{fig2}.
}
\label{fig6}
\end{figure*}

Considering the system same as that discussed in subsection $\mathrm{A}$ coupled with double electrodes under very small time-dependent voltage, then current-bias linear-response spectrum presents in \Fig{fig6} (a)-(l), in which, \Fig{fig6} (a), (c), (e), (g), (i) and (k) are linear-response function $\chi_{\rm L \rm L}(\omega)$, $\chi_{\rm L \rm R}(\omega)$, $\chi_{\rm R \rm R}(\omega)$, $\chi_{\rm R \rm L}(\omega)$, $G_{\rm L}(\omega)$ and $G_{\rm R}(\omega)$ respectively for the case of HEOM 1st-tier truncation, \Fig{fig6} (b), (d), (f), (h), (j) and (l) are the corresponding results of HEOM 2nd-tier truncation. Black (Red) solid line describes the real part of linear-response function for $\Gamma_{\rm L}=\Gamma_{\rm R}=0.2\Omega$ ($\Gamma_{\rm L}=4\Gamma_{\rm R}=0.2\Omega$) and black (red) dash line describes the corresponding imaginary part. HEOM 1st-tier truncation does not consider
energy-broadening resulting from system-electrode coupling, so in low temperature, the signals in linear-response spectrum are very sharp.
The transition from vacant-occupancy direct-product state $|0,m_{0}\rangle$ to electron single-occupancy direct-product state $|\uparrow(\downarrow),m_{0}\rangle$, $|\uparrow(\downarrow),m_{0}+1\rangle$ and
$|\uparrow(\downarrow),m_{0}+2\rangle$ results in the excitation energy $\delta\omega_{1}=0.9\Omega$, $\delta\omega_{2}=\delta\omega_{1}+1\Omega=1.9\Omega$ and $\delta\omega_{3}=\delta\omega_{2}+1\Omega=2.9\Omega$ respectively, then the transition from
electron single-occupancy direct-product state $|\uparrow(\downarrow),m_{s}\rangle$
to electron double-occupancy direct-product state $|\uparrow\downarrow,m_{s}\rangle$, $|\uparrow\downarrow,m_{s}+1\rangle$ and $|\uparrow\downarrow,m_{s}+2\rangle$ results in
the excitation energy $|\delta\omega_{1^{'}}|=|\delta\omega_{1}+U|=1.1\Omega$, $\delta\omega_{2^{'}}=|\delta\omega_{1^{'}}|+1\Omega=2.1\Omega$ and $\delta\omega_{3^{'}}=\delta\omega_{2^{'}}+1\Omega=3.1\Omega$ respectively.
For symmetric coupling, $\chi_{\rm L\rm L}(\omega)=\chi_{\rm R\rm R}(\omega)$ and $\chi_{\rm L\rm R}(\omega)=\chi_{\rm R\rm L}(\omega)$, which results in
$G_{\rm L}(\omega)=-G_{\rm R}(\omega)$. For asymmetric coupling, $\mathrm{Re(Im)}\chi_{\rm L\rm L}(\omega)>\mathrm{Re(Im)}\chi_{\rm R\rm R}(\omega)$ and $\chi_{\rm L\rm R}(\omega)=\chi_{\rm R\rm L}(\omega)$, which results in $\mathrm{Re(Im)}G_{\rm L}(\omega)>-\mathrm{Re(Im)}G_{\rm R}(\omega)$. Energy-broadening resulting from system-electrode coupling is considered enough in
HEOM 2nd-tier truncation, which induces the broadening of signals in linear-response function so that the signals of single-occupancy and that of double-occupancy are almost merged. Compared to HEOM 1st-tier truncation, additional pair-transition signal locating at $\omega=0.2\Omega$ presents in the linear-response spectrum of HEOM 2nd-tier truncation, which can account for the excitation energy $|\delta\omega|=0.2\Omega$ from electron vacant-occupancy direct-product state $|0,m_{0}\rangle$ to electron double-occupancy direct-product state $|\uparrow\downarrow,m_{0}\rangle$.

In linear-response region, a classical circuit consisting of a resistor ($R_{\rm C}$)-capacitor ($\rm C$) branch and a resistor ($R_{\rm L}$)-inductor ($\rm L$) branch connected in parallel \cite{Mo09355301} is used to
account for the dynamical admittance $G(\omega)$ of quantum elastic transport, where, $R_{\rm C}$, $R_{\rm L}$, $\rm L$ and $\rm C$ are all determined by the intrinsic
system properties described by some basis physical constants. Especially in the low frequency domain, $G(\omega)=\frac{1}{R_{\rm L}}+\omega^{2}(R_{\rm C}\rm C^{2}-
\frac{\rm L^{2}}{R^{3}_{\rm L}})$, which explains that the low frequency behavior of $G(\omega)$ is directly determined by the contest between capacity and inductor. However with strong e-ph coupling considered, the behavior of admittance can not be explained by the simple classical circuit mentioned above because of effective e-e interaction.

Besides current-bias linear-response spectrum, charge-bias and charge-gate linear-response spectrum can also characterize the information of system's excitation
and using them, so-called electrochemical
capacitance $C_{\alpha} =
\mathrm{Re}{\chi_{\rm Q\rm \alpha}(\omega = 0)}$ and $C_{g} = -\mathrm{Re}{\chi_{\rm Q\rm Q}(\omega = 0)}$ are defined to characterize
the static charging of impurity due to the bias of lead-$\alpha$ and gate voltage respectively.

\begin{figure}
\includegraphics[width=1.0\columnwidth]{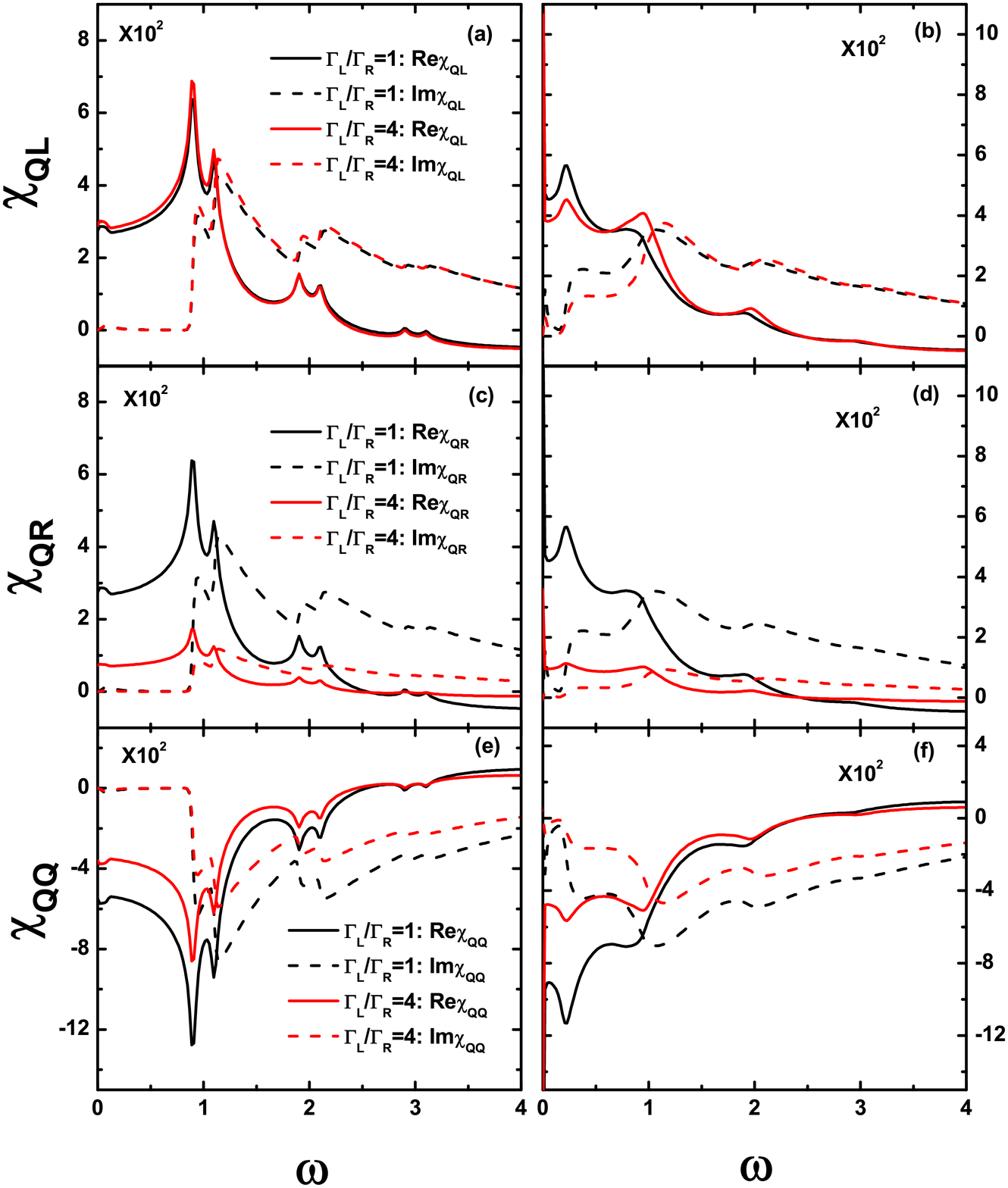}
\caption{The charge-bias and charge-gate linear-response spectrum
(in unit of
${\frac{e^{2}}{\Omega}}$),
as function of $\omega$
(in unit of $\Omega$),
for the phonon-coupled Anderson impurity QD system, with ${\varepsilon}_{s}=0.9\Omega$,
$U=-2\Omega$:
(a) $\chi_{\rm Q \rm L}(\omega)$, (c) $\chi_{\rm Q \rm R}(\omega)$ and (e) $\chi_{\rm Q \rm Q}(\omega)$
for HEOM 1st-tier truncation;
(b), (d) and (f) for HEOM 2nd-tier truncation.
The black (red) solid line means
the real part of linear-response function for symmetric (asymmetric) system-electrode coupling: $\Gamma_{\rm L}=\Gamma_{\rm R}=0.2\Omega$ ($\Gamma_{\rm L}=4\Gamma_{\rm R}=0.2\Omega$), while the black (red) dash line means the
imaginary part of linear-response function for symmetric (asymmetric) system-electrode coupling.
Other parameters are the same as those adopted in \Fig{fig2}.
}
\label{fig7}
\end{figure}

For the same system investigated in \Fig{fig6}, charge-bias and charge-gate linear-response spectrum present in \Fig{fig7} (a)-(f), in which, \Fig{fig7} (a), (c) and (e) are linear-response function $\chi_{\rm Q \rm L}(\omega)$, $\chi_{\rm Q \rm R}(\omega)$, and $\chi_{\rm Q \rm Q}(\omega)$ respectively for the case of HEOM 1st-tier truncation, \Fig{fig7} (b), (d) and (f) are the corresponding results of HEOM 2nd-tier truncation. Both for HEOM 1st-tier truncation and for HEOM 2nd-tier truncation, $C_{\rm L}$ ($C_{\rm R}$) for symmetric coupling is smaller (larger) than that for asymmetric one because $\Gamma_{\rm R}$ for symmetric coupling is larger than that for asymmetric one and $C_{g}$ for symmetric coupling is larger than that for asymmetric one because of the same reason.
But obviously, 1st-tier truncation underestimates the values of $C_{\alpha}$ and $C_{g}$ and the curves have the reverse tendency near $\omega=0$ to that of 2nd-tier truncation.
The signals of inelastic excitation are also presented in HEOM 1st-tier and 2nd-tier truncation.
Compared to HEOM 1st-tier truncation, the charge-bias and charge-gate linear-response of HEOM 2nd-tier truncation present the wider broadening and the signals of elastic pair-transition excitation locating at $\omega=0.2\Omega$ appear.

Generally, for linear-response spectrum of arbitrary type, we have the symmetries: $\mathrm{Re}\chi(\omega)=\mathrm{Re}\chi(-\omega)$ and $\mathrm{Im}\chi(\omega)=-\mathrm{Im}\chi(-\omega)$.

\section{Concluding remarks}
\label{thsec4}
We have combined the HEOM formalism with
the small polaron transformation to study quantum transport through impurity QD systems with strong e-ph coupling in charge-Kondo regime.
The Einstein lattice model is adopted to describe phonon bath. HEOM combined with small polaron transformation is a nonpertubative tool to
study the steady and time-dependent transport, with
a broad range of couplings beyond the Einstein lattice model.

Besides the self-energy correction, a strong e-ph coupling will result in the effective e-e attraction. The physical process of negative-$U$ transport model can be revealed by mapping it to positive-$U$ model by particle-hole/left-right transformation. In this transformation, when a symmetric bias is applied,, it can be understood by a positive-$U$ charge-battery model, however, when an asymmetric bias is applied, it can be understood by so called spin-cell model. All the signals presented in DOS can be classified into sequential tunneling, pair-transition and inelastic cotunneling processes. Under the symmetric bias, the signals of pair-transition and inelastic cotunneling present the normal non-equilibrium Kondo split, while under the
asymmetric bias, the signals of pair-transition present the abnormal split.
Differential conductance spectrum under the asymmetric bias presents the corresponding signals, which can be modulated by the system-electrode coupling and bath-temperature.
Enhancement of bath-temperature strengthens the phonon-absorption (emission)--assisted sequential tunneling and the inelastic pair-transition process.
As a supplement to steady transport, inelastic dynamics driven by ramp-up time-dependent voltage presents clear steps tailored by the duration-time and bath-temperature and linear-response spectrums including current-bias spectrum, charge-bias spectrum and charge-gate one reveal the signals of electrons' dynamical transition.

Noticing that in the traditional superconductor, e-ph interaction also makes Cooper-pair (two electrons with reverse momentum and spin), which is similar to the
double occupation of QD caused by strong e-ph coupling, so the idea of considering the steady and transient transport of superconductor--normal(superconducting)-QD--normal-metal (superconductor) hybrid system is straightforward. It is expected that the strong e-ph coupling may play a vital role in Andreev-reflection or Josephson effect, which is a quite interesting topic. The corresponding work is in process.

\acknowledgments
    Support from the University Grants Committee
of Hong Kong SAR (AoE/P-04/08-2) and NSFC under Grants No.11204180
is gratefully acknowledged.

\appendix

\section{Transformation from the negative-\emph{U} model
  to an equivalent positive-\emph{U} one}
\label{thapp_phlr}

\subsection{Particle-hole/left-right transformation}
Let us carry out particle-hole/left-right transformation for the inelastic quantum transport model based on small polaron transformation as follows:
\begin{align}
\hat{\beta}_{L(R),k,\downarrow}&=\hat{c}^{\dag}_{R(L),-k,\downarrow},\nl
\hat{\beta}_{a_{\downarrow}}&=-\hat{a}^{\dag}_{\downarrow}.
\end{align}

We can conclude from the transformation above that $f(\tilde{\epsilon}_{L(R) k\downarrow})=f(\epsilon_{R(L) k\downarrow})$ and the spin-down single-occupation state in central region $|\tilde{\downarrow}\rangle$ is equivalent to the vacant-occupation state in Fock space $|0\rangle$ with the notation $\tilde{.}$ denoting the hole.

Noting that $\beta$ operator satisfies anti-communication relations of Fermion, then, the total hamiltonian can be rewritten as follows:
\begin{widetext}
\begin{align}
H_{\rm T}&=H_{\rm res}+H_{\rm e}+H_{\rm e-res}+H_{\rm ph},\nl
H_{\rm res}&=\sum_{\alpha,k}(\epsilon_{\alpha,k,\uparrow}-eV_{\alpha})\hat{n}_{\alpha,k,\uparrow}
+\sum_{k}(\epsilon_{L,k,\downarrow}-eV_{L})(1-\hat{\tilde{n}}_{R,-k,\downarrow})
+(\epsilon_{R,k,\downarrow}-eV_{R})(1-\hat{\tilde{n}}_{L,-k,\downarrow}),\nl
H_{\rm e}&=(\varepsilon_{\uparrow}+U)\hat{n}_{\uparrow}+\varepsilon_{\downarrow}(1-\hat{\tilde{n}}_{\downarrow})
-U\hat{n}_{\uparrow}\hat{\tilde{n}}_{a_{\downarrow}},\nl
H_{\rm e-res}&=\sum_{\alpha,k}(t_{\alpha,k,\uparrow}\hat{c}^{\dag}_{\alpha,k,\uparrow}\hat{a}_{\uparrow}e^{i\hat{\varphi}}+{\rm H.c.})
+(t_{L,k,\downarrow}\hat{\beta}^{\dag}_{a_{\downarrow}}\hat{\beta}_{R,-k,\downarrow}e^{i\hat{\varphi}}+{\rm
H.c.})+(t_{R,k,\downarrow}\hat{\beta}^{\dag}_{a_{\downarrow}}\hat{\beta}_{L,-k,\downarrow}e^{i\hat{\varphi}}+{\rm
H.c.}),\nl
H_{\rm ph}&=\sum_{q}\omega_{q}\hat{d}^{\dag}_{q}\hat{d}_{q},
\end{align}
where, $\hat{\tilde{n}}_{\alpha,k,\downarrow}=\hat{\beta}^{\dag}_{\alpha,k,\downarrow}\hat{\beta}_{\alpha,k,\downarrow}$ and $\hat{\tilde{n}}_{a_{\downarrow}}
=\hat{\beta}^{\dag}_{a_{\downarrow}}\hat{\beta}_{a_{\downarrow}}$.

Noting the appointment and abandoning the irrelevant additive constant, $\tilde{H}_{\rm T}=H_{\rm T}+c$ can be expressed as follows:
\begin{align}
\tilde{H}_{\rm T}&=\tilde{H}_{\rm res}+\tilde{H}_{\rm e}+\tilde{H}_{\rm e-res}+\tilde{H}_{\rm ph},\nl
\tilde{H}_{\rm res}&=\sum_{k}(\epsilon_{L,k,\uparrow}-eaV)\hat{\tilde{n}}_{L,k,\uparrow}+\sum_{k}(\epsilon_{L,k,\downarrow}+e(a-1)V)\hat{\tilde{n}}_{L,k,\downarrow}
+\sum_{k}(\epsilon_{R,k,\uparrow}-e(a-1)V)\hat{\tilde{n}}_{R,k,\uparrow}+\sum_{k}(\epsilon_{R,k,\downarrow}+eaV)\hat{\tilde{n}}_{R,k,\downarrow},\nl
\tilde{H}_{\rm e}&=(\varepsilon_{\uparrow}+U)\hat{\tilde{n}}_{a_{\uparrow}}+(-\varepsilon_{\downarrow})\hat{\tilde{n}}_{a_{\downarrow}}-U\hat{\tilde{n}}_{a_{\uparrow}}\hat{\tilde{n}}_{a_{\downarrow}},\nl
\tilde{H}_{\rm e-s}&=\sum_{\alpha,k,\sigma}\tilde{t}_{\alpha,k,\sigma}\hat{\tilde{c}}^{\dag}_{\alpha,k,\sigma}\hat{\tilde{a}}_{\sigma}e^{i\sigma\hat{\varphi}}+{\rm H.c.},\nl
\tilde{H}_{\rm ph}&=\sum_{q}\omega_{q}\hat{d}^{\dag}_{q}\hat{d}_{q},
\end{align}
\end{widetext}
where, $\hat{\tilde{n}}_{\alpha,k,\uparrow}=\hat{n}_{L,k,\uparrow}$, $\hat{\tilde{n}}_{L(R),k,\downarrow}=1-\hat{n}_{R(L),-k,\downarrow}$, $\hat{\tilde{n}}_{a_{\uparrow}}=\hat{n}_{\uparrow}$, $\hat{\tilde{n}}_{a_{\downarrow}}=1-\hat{n}_{\downarrow}$, $\hat{\tilde{c}}_{\alpha,k,\uparrow}=\hat{c}_{\alpha,k,\uparrow}$, $\hat{\tilde{c}}_{L(R),k,\downarrow}=\hat{c}^{\dag}_{R(L),-k,\downarrow}$,
$\hat{\tilde{a}}_{\uparrow}=\hat{a}_{\uparrow}$, $\hat{\tilde{a}}_{\downarrow}=-\hat{a}^{\dag}_{\downarrow}$,
$\tilde{t}_{\alpha,k,\uparrow}=t_{\alpha,k,\uparrow}$ and $\tilde{t}_{L(R),k,\downarrow}=t_{R(L),k,\downarrow}$,
$\sigma$ in exponent denotes $+1$ for spin up or $-1$ for spin down.

After particle-hole/left-right transformation is implemented, the current relation between positive-\emph{U} model and negative-\emph{U} model is as follows:
\begin{align}
\tilde{I}_{\alpha\uparrow}&=-\langle\frac{d}{dt}\sum_{k}\hat{\tilde{n}}_{\alpha,k,\uparrow}\rangle=-\langle\frac{d}{dt}\sum_{k}\hat{n}_{\alpha,k,\uparrow}\rangle=I_{\alpha\uparrow},\nl
\tilde{I}_{L(R)\downarrow}&=-\langle\frac{d}{dt}\sum_{k}\hat{\tilde{n}}_{L(R),k,\downarrow}\rangle=\langle\frac{d}{dt}\sum_{k}\hat{n}_{R(L),k,\downarrow}\rangle=-I_{R(L)\downarrow}.
\end{align}

The particle number relation between positive-\emph{U} model and negative-\emph{U} model is as follows:
\begin{align}
\tilde{N}_{\uparrow}&=\langle \hat{\tilde{n}}_{a_{\uparrow}}\rangle=\langle \hat{n}_{\uparrow}\rangle=N_{\uparrow},\nl
\tilde{N}_{\downarrow}&=\langle \hat{\tilde{n}}_{a_{\downarrow}}\rangle=1-\langle \hat{n}_{\downarrow}\rangle=1-N_{\downarrow}.
\end{align}

The population relation between positive-\emph{U} model and negative-\emph{U} model is as follows:
\begin{align}
\tilde{P}_{0}&=P_{\downarrow},\nl
\tilde{P}_{\uparrow}&=P_{d},\nl
\tilde{P}_{\downarrow}&=P_{0},\nl
\tilde{P}_{d}&=P_{\uparrow},
\end{align}
where, $\tilde{P}_{0}(P_{0})$, $\tilde{P}_{\uparrow}(P_{\uparrow})$, $\tilde{P}_{\downarrow}(P_{\downarrow})$, $\tilde{P}_{d}(P_{d})$ are vacant-occupancy, spin-up single-occupancy, spin-down single-occupancy, double-occupancy after transformation (before transformation) respectively.

The density of states after transformation can be expressed as follows:
\begin{align}
\tilde{A}_{\uparrow}(\omega)&=\frac{1}{\pi}\mathrm{Re}[\int^{\infty}_{0}dt\langle\{\hat{\tilde{a}}_{\uparrow}(t),\hat{\tilde{a}}^{\dag}_{\uparrow}\}\rangle e^{i\omega t}]=A_{\uparrow}(\omega),\nl
\tilde{A}_{\downarrow}(\omega)&=\frac{1}{\pi}\mathrm{Re}[\int^{\infty}_{0}dt\langle\{\hat{\tilde{a}}_{\downarrow}(t),\hat{\tilde{a}}^{\dag}_{\downarrow}\}\rangle e^{i\omega t}]\nl
&=\frac{1}{\pi}\mathrm{Re}[\int^{\infty}_{0}dt\langle\{\hat{a}_{\downarrow}(t),\hat{a}^{\dag}_{\downarrow}\}\rangle e^{-i\omega t}]=A_{\downarrow}(-\omega).
\end{align}

Since $\tilde{A}_{\uparrow}(\omega)=A_{\uparrow}(\omega)$ and $\tilde{A}_{\downarrow}(\omega)=A_{\downarrow}(-\omega)$, in equilibrium, the charge-Kondo peak
locating at $\omega=\varepsilon_{\uparrow}+\varepsilon_{\downarrow}+U$ is robust under magnetic field, while spin-Kondo peak locating at $\omega=0$ will generate Zeeman split.

\subsection{Symmetric bias versus asymmetric bias}
When symmetric bias is applied, $a=1/2$, then the total hamiltonian can be expressed as an ordinary form:
\begin{align}
\tilde{H}_{\rm T}&=\tilde{H}_{\rm res}+\tilde{H}_{\rm e}+\tilde{H}_{\rm e-res}+\tilde{H}_{\rm ph},\nl
\tilde{H}_{\rm res}&=\sum_{k\sigma}(\epsilon_{L,k,\sigma}-\frac{1}{2}eV)\hat{\tilde{n}}_{L,k,\sigma}
+\sum_{k\sigma}(\epsilon_{R,k,\sigma}+\frac{1}{2}eV)\hat{\tilde{n}}_{R,k,\sigma},\nl
\tilde{H}_{\rm e}&=(\varepsilon_{\uparrow}+U)\hat{\tilde{n}}_{a_{\uparrow}}+(-\varepsilon_{\downarrow})\hat{\tilde{n}}_{a_{\downarrow}}+(-U)\hat{\tilde{n}}_{a_{\uparrow}}\hat{\tilde{n}}_{a_{\downarrow}},\nl
\tilde{H}_{\rm e-s}&=\sum_{\alpha,k,\sigma}\tilde{t}_{\alpha,k,\sigma}\hat{\tilde{c}}^{\dag}_{\alpha,k,\sigma}\hat{\tilde{a}}_{\sigma}e^{i\sigma\hat{\varphi}}+{\rm H.c.},\nl
\tilde{H}_{\rm ph}&=\sum_{q}\omega_{q}\hat{d}^{\dag}_{q}\hat{d}_{q}.
\end{align}

For the positive-$U$ model transformed from negative-$U$ model, the spectrum density can be expressed as follows:
\begin{align}
\tilde{J}_{L(R)\uparrow}(\omega)&=\sum_{k}|\tilde{t}_{L(R),k,\uparrow}|^{2}\delta(\omega\pm\frac{1}{2}eV-\epsilon_{L(R),k,\uparrow})\nl
&=\sum_{k}|t_{L(R),k,\uparrow}|^{2}\delta(\omega\pm\frac{1}{2}eV-\epsilon_{L(R),k,\uparrow}),\nl
\tilde{J}_{L(R)\downarrow}(\omega)&=\sum_{k}|\tilde{t}_{L(R),k,\downarrow}|^{2}\delta(\omega\pm\frac{1}{2}eV-\epsilon_{L(R),k,\downarrow})\nl
&=\sum_{k}|t_{R(L),k,\downarrow}|^{2}\delta(\omega\pm\frac{1}{2}eV-\epsilon_{L(R),k,\downarrow}).
\end{align}

Specially, when symmetric system-bath coupling is considered, i.e. $t_{L,k,\sigma}=t_{R,k,\sigma}=t_{k,\sigma}$, and symmetric level $\varepsilon_{\sigma}=-U/2$ is considered, the negative-$U$ model based on small polaron transformation is converted to positive-$U$ model as follows:
\begin{align}
\tilde{H}_{\rm T}&=\tilde{H}_{\rm res}+\tilde{H}_{\rm e}+\tilde{H}_{\rm e-res}+\tilde{H}_{\rm ph},\nl
\tilde{H}_{\rm res}&=\sum_{k\sigma}(\epsilon_{L,k,\sigma}-\frac{1}{2}eV)\hat{\tilde{n}}_{L,k,\sigma}
+\sum_{k\sigma}(\epsilon_{R,k,\sigma}+\frac{1}{2}eV)\hat{\tilde{n}}_{R,k,\sigma},\nl
\tilde{H}_{\rm e}&=\frac{U}{2}\sum_{\sigma}\hat{\tilde{n}}_{a_{\sigma}}+(-U)\hat{\tilde{n}}_{a_{\uparrow}}\hat{\tilde{n}}_{a_{\downarrow}},\nl
\tilde{H}_{\rm e-res}&=\sum_{\alpha,k,\sigma}t_{k,\sigma}\hat{\tilde{c}}^{\dag}_{\alpha,k,\sigma}\hat{\tilde{a}}_{\sigma}e^{i\sigma\hat{\varphi}}+{\rm H.c.},\nl
\tilde{H}_{\rm ph}&=\sum_{q}\omega_{q}\hat{d}^{\dag}_{q}\hat{d}_{q}.
\end{align}

Compared with the traditional positive-\emph{U} model based on small polaron transformation, the only difference is that $e^{i\hat{\varphi}}$ converts to $e^{i\sigma\hat{\varphi}}$, which equivalents to implementing the unitary transformation $\hat{X}=e^{-i\sum_{\sigma}\sigma\hat{n}_{\sigma}\hat{\varphi}}$ to the e-ph coupling system with $H_{\rm e-\rm ph}=\sum_{q}g_{q}(\hat{d}^{\dag}_{q}
+\hat{d}_{q})\sum_{\sigma}\sigma\hat{n}_{\sigma}$. The difference between $e^{i\hat{\varphi}}$ and $e^{i\sigma\hat{\varphi}}$ has no impaction to the final physical quantities because phonon correlation function $C^{\sigma}_{\rm ph}(t)=\sum^{+\infty}_{m=-\infty}A_{m}e^{-im\Omega t}$ is irrelevant to $\sigma$, the same HEOM can be constructed. \cite{Jia12245427}

When asymmetric bias is applied, $a=1$, then the total hamiltonian can be expressed as follows:
\begin{align}
\tilde{H}_{\rm T}&=\tilde{H}_{\rm res}+\tilde{H}_{\rm e}+\tilde{H}_{\rm e-res}+\tilde{H}_{\rm ph},\nl
\tilde{H}_{\rm res}&=\sum_{k}(\epsilon_{L,k,\uparrow}-eV)\hat{\tilde{n}}_{L,k,\uparrow}+\sum_{k}(\tilde{\epsilon}_{L,k,\downarrow}-eV)\hat{\tilde{n}}_{L,k,\downarrow}\nl
&+\sum_{k}\epsilon_{R,k,\uparrow}\hat{\tilde{n}}_{R,k,\uparrow}+\sum_{k}\tilde{\epsilon}_{R,k,\downarrow}\hat{\tilde{n}}_{R,k,\downarrow},\nl
\tilde{H}_{\rm e}&=(\varepsilon_{\uparrow}+U)\hat{\tilde{n}}_{a_{\uparrow}}+(-\varepsilon_{\downarrow})\hat{\tilde{n}}_{a_{\downarrow}}-U\hat{\tilde{n}}_{a_{\uparrow}}\hat{\tilde{n}}_{a_{\downarrow}},\nl
\tilde{H}_{\rm e-res}&=\sum_{\alpha,k,\sigma}\tilde{t}_{\alpha,k,\sigma}\hat{\tilde{c}}^{\dag}_{\alpha,k,\sigma}\hat{\tilde{a}}_{\sigma}e^{i\sigma\hat{\varphi}}+{\rm H.c.},\nl
\tilde{H}_{\rm ph}&=\sum_{q}\omega_{q}\hat{d}^{\dag}_{q}\hat{d}_{q}.
\end{align}
where, $\tilde{\epsilon}_{L(R),k,\downarrow}=\epsilon_{L(R),k,\downarrow}+eV$.

The corresponding spectrum density can be expressed as follows:
\begin{align}
\tilde{J}_{L\uparrow}(\omega)&=\sum_{k}|\tilde{t}_{L,k,\uparrow}|^{2}\delta(\omega+eV-\epsilon_{L,k,\uparrow})\nl
&=\sum_{k}|t_{L,k,\uparrow}|^{2}\delta(\omega+eV-\epsilon_{L,k,\uparrow}),\nl
\tilde{J}_{L\downarrow}(\omega)&=\sum_{k}|\tilde{t}_{L,k,\downarrow}|^{2}\delta(\omega-\epsilon_{L,k,\downarrow})=\sum_{k}|t_{R,k,\downarrow}|^{2}\delta(\omega-\epsilon_{L,k,\downarrow}),\nl
\tilde{J}_{R\uparrow}(\omega)&=\sum_{k}|\tilde{t}_{R,k,\uparrow}|^{2}\delta(\omega-\epsilon_{R,k,\uparrow})=\sum_{k}|t_{R,k,\uparrow}|^{2}\delta(\omega-\epsilon_{R,k,\uparrow})\nl
\tilde{J}_{R\downarrow}(\omega)&=\sum_{k}|\tilde{t}_{R,k,\downarrow}|^{2}\delta(\omega-eV-\epsilon_{R,k,\downarrow}),\nl
&=\sum_{k}|t_{L,k,\downarrow}|^{2}\delta(\omega-eV-\epsilon_{R,k,\downarrow}).
\end{align}

So, the problem can be transformed into the model of spin battery in which, a positive-$U$ quantum dot coupled to electrodes which has different Fermi level with different spin, if Lorentz spectrum is adopted, the spectrum density expressions are as follows:
\begin{align}
\tilde{J}_{L\uparrow}(\omega)&=\frac{1}{2\pi}\frac{\Gamma W^{2}}{(\omega-E^{\uparrow}_{f}+eV)^{2}+W^{2}},\nl
\tilde{J}_{L\downarrow}(\omega)&=\frac{1}{2\pi}\frac{\Gamma W^{2}}{(\omega-E^{\downarrow}_{f}+eV)^{2}+W^{2}},\nl
\tilde{J}_{R\uparrow}(\omega)&=\frac{1}{2\pi}\frac{\Gamma W^{2}}{(\omega-E^{\uparrow}_{f})^{2}+W^{2}},\nl
\tilde{J}_{R\downarrow}(\omega)&=\frac{1}{2\pi}\frac{\Gamma W^{2}}{(\omega-E^{\downarrow}_{f})^{2}+W^{2}},
\end{align}
where, $E^{\uparrow}_{f}=0$ and $E^{\downarrow}_{f}=eV$ are the Fermi level of spin up and spin down respectively, $\Gamma$ is the coupling between system and the electrode, $W$ is the bandwidth of the electrode.

\bibliographystyle{aip}
\bibliography{bibrefs}

\end{document}